\documentclass[useAMS,usenatbib,usegraphicx]{mn2e}
\usepackage{times}

\title[3 $-$ 200 $\mu$m emission in an isolated local translucent
cloud]{{\it ISO} observations of 3 $-$ 200 $\mu$m
emission by three dust populations in an isolated local translucent cloud\thanks{Based
on observations made with {\it ISO}, an ESA project with instruments funded by ESA member
states (especially the PI countries: France, Germany, the Netherlands and the United
Kingdom) and with the participation of ISAS and NASA.}}

\author[M. G. Rawlings et al.]{M. G.
Rawlings$^{1}$\thanks{E-mail: mark@astro.helsinki.fi},
M. Juvela$^{1}$, K. Mattila$^{1}$, K. Lehtinen$^{1}$ \& D. Lemke$^{2}$\\
$^{1}$Observatory, P.O. Box 14, University of Helsinki, Helsinki, FIN-00014, Finland\\
$^{2}$Max-Planck-Institut f\"ur Astronomie, K\"onigstuhl 17, Heidelberg, D-69117, Germany}
\begin{document}

\date{Accepted 2004 October 21. Received 2004 October 18; in original form 2003 November 28}

\pagerange{\pageref{firstpage}--\pageref{lastpage}} \pubyear{2004}

\maketitle

\label{firstpage}

\begin{abstract}
We present ISOPHOT spectrophotometry of three positions within the isolated high
latitude cirrus cloud G~300.2~$-$~16.8, spanning from the near- to far-infrared. The positions
exhibit contrasting emission spectrum contributions from the UIBs, very small grains
and large classical grains, and both semi-empirical and numerical models are
presented. At all three positions, the UIB spectrum shapes are found to be similar, and
the large grain emission may be fitted by an equilibrium temperature of $\sim$17.5 K. The energy
requirements of both the observed emission spectrum and optical scattered light are
shown to be satisfied by the incident local ISRF. The FIR emissivity of dust in
G~300.2~$-$~16.8 is found to be lower than in globules or dense clouds, and is
even lower than model predictions for dust in the diffuse ISM. The results suggest
physical differences in the ISM mixtures between positions within the cloud, possibly arising from
grain coagulation processes.
\end{abstract}


\section{Introduction}
\subsection{UIBs and G~300.2~$-$~16.8}
A wide range of Galactic astrophysical objects exhibit infrared (IR) emission spectra
clearly arising from multiple components in the solid- and gas-phases. Emission in the
so-called Unidentified Infrared Bands (UIBs or UIR bands, also termed Infrared
Emission Features, or IEFs) is significant in a wide range of sources. It often makes
up as much as $20 - 30$ per cent of the total IR emission, and is thought to arise due to the
presence of carbonaceous material of some kind.
Although solid-phase carbonaceous grain components
have been suggested, e.g. hydrogenated amorphous hydrocarbon (HAC;
\citealt{jones}), quenched carbonaceous composites (QCC;
\citealt{sakata}) or coal \citep{papoular}, the
bands are still widely attributed to gas-phase material.
Typically, large free-floating aromatic species such as
Polycyclic Aromatic Hydrocarbons (PAHs) are proposed (\citealt{leger};
\citealt{allamandola}),
and are usually thought to be ionized or to feature modified hydrogenation
states. PAHs are excited by incident UV/visual radiation and subsequently re-emit in
the IR at the specific UIB wavelengths. Observations have revealed that the bands are
ubiquitous, appearing towards a range of regions such as Planetary Nebulae (PNe), HII
regions and reflection nebulae around early-type stars. The presence of the 3.3- and
6.2-$\mu$m bands was mapped by AROME observations of the diffuse Galactic disc
emission (\citealt{giard}; \citealt{ristorcelli}) and the full spectrum between 5
and 11.5 $\mu$m by ISOPHOT and IRTS spectrophotometry towards the inner
Galaxy (\citealt{mattila96}; \citealt{tanaka}; \citealt{onaka}).
\citet{kahanpaa} recently observed the inner Galaxy
($\mid l\mid \le 60^{\circ}$, $\mid
b\mid \le 1^{\circ}$) UIB spectrum from $5 - 11.5$ $\mu$m along 49 sightlines
using ISOPHOT data. IRTS results for several areas at $\mid b\mid \le 4^{\circ}$ both in the inner and outer Galaxy
have been presented by \citet{sakon}. The band strengths have also been mapped along the disc of an
edge-on spiral galaxy (NGC 891) similar to the Milky Way \citep{mattila99}, and
subsequently in other spiral galaxies \citep{lu}.

The first discovery and measurement of the UIB emission in high latitude cirrus
clouds was via their broadband {\it IRAS} 12-$\mu$m detections (\citealt{low};
\citealt{boulanger85}; see \citealt{verter} and references therein). However,
the first spectrally-resolved measurements of the bands toward an individual isolated
cirrus cloud (G~300.2~$-$~16.8) externally heated by the interstellar radiation field
(ISRF) were made by \citet[][ hereinafter L98]{lemke98} using the ISOPHOT
instrument \citep{lemke96} on board ESA's {\it ISO} spacecraft
\citep{kessler}.

The cloud G~300.2~$-$~16.8 is thought to be at the same distance of $150 \pm 20$ pc from the Sun as the Chamaeleon clouds in general \citep{knude}. Situated at $z \sim -40$ pc below the Galactic plane, it is well within the half-thickness of the local Galactic dust layer (see e.g. \citealt{neckel}).
\citet{bernard93} estimated G~300.2~$-$~16.8 to have a diameter of $\sim$ 3 pc and a mass of 60 $M_{\sun}$. {\it IRAS} photometry indicated that the emission at 12, 25, 60 and 100 $\mu$m peaks at different positions within the cloud, suggesting that the object could be used to probe potential differences in the dust composition and the ISRF.

The L98 $3.3-16$ $\mu$m observations of
G~300.2~$-$~16.8 determined that the UIBs toward this object had absolute intensities
$\sim$1/1000th of those typically seen toward bright planetary or reflection nebulae, but
with relative band intensities not markedly different from these environments. This
claim of a similarly-shaped UIB spectrum was confirmed by ISOCAM-CVF spectroscopy
\citep{boulanger98}. In addition, they reported a non-zero continuum level at 10 and
16 $\mu$m, which they attributed to a population of very small grains (VSGs).
\subsection{Three-component model}
A three-component IR dust model has been proposed by e.g. \citet{puget}.
The first of these components is a mixture of large aromatic organic ions or molecules (e.g.
compact PAHs) of $10 - 1000$ atoms (\citealt{bakes} \& references therein), that is
thought to give rise to the UIBs. The second is a population of transiently-heated VSGs
\citep{sellgren} producing non-thermal emission in the
mid- to far-IR. These are widely believed to be carbonaceous in nature, despite
tight cosmic abundance constraints on carbon (\citealt{snow};
\citealt{kim}). The third population is of larger ($\sim100 - 2000$ \AA ) `classical'
dust grains emitting in thermal equilibrium in the far-IR ($\lambda \geq 80 \mu$m) at normal
ISM temperatures of $\le$ 20K. This population corresponds to the large grains used in
popular models such as those of \citet{mathis96} and
\citet{li97}, and may have a fluffy and/or core-mantle structure.
\subsection{Comparison of UIB and dust emission}
To date, however, there have been no comprehensive comparisons of the near- to far-
IR emission spectra of these UIB-rich environments over a wide range of UV/optical
field strengths. The acquisition of a uniform dataset in order to thoroughly investigate
ISM models of this kind is therefore needed. The {\it IRAS} dataset only features four
broad photometric bands (centred at 12, 25, 60 and 100 $\mu$m) and is consequently
of limited use in terms of spectral coverage. Data obtained using DIRBE suffers from
excessively coarse spatial resolution (0.7$^{\circ}$). Furthermore, any additional FIR
data subsequently acquired (e.g. via airborne experiments) have generally been
piecemeal in nature. This situation changed with the advent of {\it ISO},
which offered an unprecedented combination of IR wavelength coverage
and detector sensitivities. We therefore begin here a series of papers that will address
the situation by presenting a more complete spectrophotometric dataset obtained using
ISOPHOT. This first paper focuses on the isolated
cirrus/translucent cloud G~300.2~$-$~16.8, and significantly augments the number of
photometric bands used, and hence the wavelength coverage, of the L98 dataset.

Using ISOPHOT-P1, P2, C100 and C200 between 3.1 and 200 $\mu$m, we observed
the cirrus/translucent cloud G~300.2~$-$~16.8 in the Chamaeleon dark cloud
complex. G~300.2~$-$~16.8 has previously been found to exhibit a
large {\it IRAS} $I_{12 \mu\rm{m}}$ /
$I_{100 \mu\rm{m}}$ ratio of 0.14 \citep{laureijs89b}.
Furthermore, the signal in the
{\it IRAS} 12-, 25-, 60- and 100-$\mu$m bands peaks at different positions in the cloud,
suggesting variations in the local dust composition. Nevertheless, ISOPHOT data
presented in L98 supported the detection of a UIB spectrum with a shape similar to
that of a high ISRF environment. The comparatively high Galactic latitude ($b = -16.8$
degrees) minimizes the risk of confusion with unrelated structures along the sightline.

In this paper, we present an analysis of the three dust components
and their variation within a single cloud. Section 2 details the observations and
our data reduction methods and Sect. 3 summarizes the results. In Sect. 4, we
use USNO-B1.0 and 2MASS archive data to derive the extinction
properties of the sightlines. Section 5 describes our first attempts to model the
emission across the broad ISOPHOT wavelength range, using both semi-empirical and
physical numerical models. In Sect. 6, we consider the energy budget of the cloud,
and we estimate FIR opacities and column densities in Sect. 7. Section 8 offers some
conclusions and discusses some of the astrophysical implications of this work, and
we conclude with a summary of the key points in Sect. 9.

\begin{figure*}
\center
\includegraphics[angle=0,width=16cm]{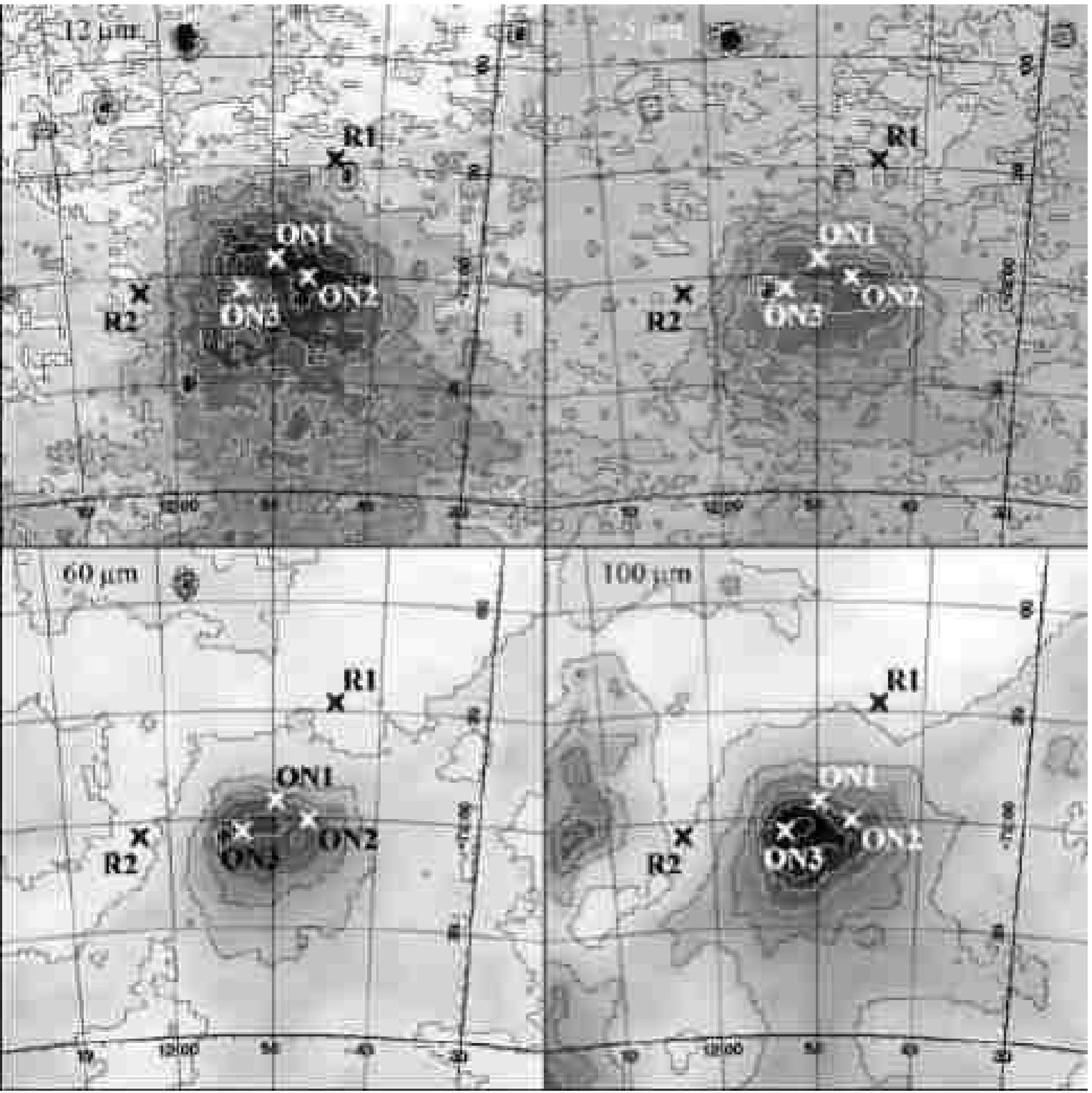}

 \caption{Infrared observations of the isolated cirrus cloud G~300.2~$-$~16.8. The images are the IRAS maps for the four IRAS filter bands. Crosses mark the position of observed by us using the ISOPHOT instrument aboard the ISO spacecraft, where R1 and R2 are the off-source reference positions and ON1, ON2 and ON3 are the on-source positions. The contours for the 12-$\mu$m filter range from 1.2 to 2.6 in steps of 0.2. The contours for the 25-$\mu$m filter range from $4.7-6.1$ in steps of 0.2. The contours for the 60-$\mu$m filter range from $1.0-7.0$ in steps of 1.0. The contours for the 100-$\mu$m filter range from $10.0-28.0$ in steps of 3.0. Units are MJy sr$^{-1}$. Co-ordinates are  $\alpha , \delta (B1950.0)$.}

\label{Fig1}

\end{figure*}

\section{Observations and reductions}

The observations were carried out during {\it ISO} revolutions 708 and 718 (1997 October 24
and 1997 November 3; for observation TDTs see Table 1). The observed on-source
and reference positions match those in L98, with the three ON positions selected to
coincide with the well-identified {\it IRAS} 12, 25 and 100-$\mu$m maxima (see Fig. 1).

Observations were performed in ten filters that are listed in Table 1, along with the
PHT-P aperture diaphragm sizes or the PHT-C camera field sizes.
The integration time was 32 seconds in all
cases. The sparse map observing templates (AOTs 17/18/19: PHT-P and 37/38/39:
PHT-C) were used \citep{klaas94} and a separate sparse map was obtained using
each filter. Since the sky brightness changes in the mid-IR bands by only
a few percent between the different ON- and REF-positions in a given map, these modes
were used to minimize detector drift effects that dominate the errors at these wavebands.
\subsection{Reduction procedure}
The data reduction was performed using the ISOPHOT Interactive Analysis Program
(PIA) Version V9.0 \citep{gabriel}. Essentially, the same reduction
steps were applied as described in L98.

\begin{table*}
  \vbox to220mm{\vfil Landscape table to go here.
  \caption{}
 \vfil}
 \label{landtable}
\end{table*}
\subsection{Calibration}
Our method utilizes the Zodiacal Light (ZL) as a calibration source. The basic
technique was already used in L98 and has been described in more detail by
\citet{acostapulido}. The first and last measurements of each sparse map were
paired with a measurement of the on-board calibration source FCS1 
\citep{lemke96}, which was heated to give a signal corresponding to the expected sky
brightness, yielding a first calibration. We then used the Zodiacal
emission values from {\it COBE}/DIRBE to generate responsivity corrections for the
resultant REF-position fluxes at wavelengths of $3 - 60 \mu$m. The 60-$\mu$m cut-off
point was chosen since the ZL no longer dominates the emission continuum longward
of this wavelength. The photometric responsivity correction factors were determined
by using the ISOPHOT ZL template spectrum ($5.9 - 11.7$ $\mu$m) of
\citet{leinert} most closely matching the G~300.2~$-$~16.8 position
(spectrum 8). The spectrum was rescaled to match the colour-corrected
monochromatic {\it COBE}/DIRBE
fluxes at 3.5, 5, 12 and 25 $\mu$m. This was
fitted using a blackbody function, which was extrapolated out
to 60 $\mu$m, and taken as the actual reference ZL levels.
Figure 2 shows the Zodiacal emission spectrum out to 25 $\mu$m, the
{\it COBE}/DIRBE measurements \citep{hauser}, and our new
photometry. The L98 data have been recalibrated using the improved
ZL templates, again accounting for position and observation times.
The derived scaling factors were applied to our
colour-corrected (ON-REF) photometry.
\subsection{Signal drift and foreground subtraction}
We display the observation sequences for the ten sparse maps in Fig. 3. In all ten of
the filter bands, there is a clear excess signal in the ON positions. The detector drift
has been modelled in one of two ways. For the cases where more than one measurement was
made at each reference position, the background emission baseline was fitted using a
function of the same
form as that used in L98, otherwise a linear fit to the reference position data points
was used. Subtraction of these fitted lines yielded the cloud emission.
Photometric data were then corrected using standard {\it ISO}
colour correction tables, to produce monochromatic fluxes directly comparable to
those presented in L98. For filters with reference wavelengths greater than
20 $\mu$m, this was done by fitting either one or two modified blackbody functions of
the form $\nu^{2} B_{\nu}$ and using the derived temperatures in conjunction with
interpolated ISOPHOT colour correction lookup tables. No colour corrections were
applied at the shorter wavelengths ($\lambda \le 16 \mu$m. For the
C100 and C200 data, the surface brightness was determined from all of the detector
array elements (3$\times$3 for C100 and 2$\times$2 for C200) in an attempt to
optimally match the apertures used for the P1 and P2 measurements and hence minimize
any possible systematic errors due to sampling area.
\subsection{Error analysis}
{\it Statistical errors} for the new data have been estimated by two methods, as in L98:
\begin{enumerate}
\item Internal errors, $\sigma_{\rm INT}$, for each measurement are obtained from the PIA
analysis, and their means are given in column 9 of Table 1;
\item External errors, $\sigma_{\rm{EXT}}$, are obtained from the comparison of two
independent measurements for O1, O2, O3 and R2 in the $10 - 25$-$\mu$m filters, and of three independent measurements of a reference position (R1 or R2) in the $7.3 - 7.7$-$\mu$m and $60 - 200$-$\mu$m filters.
\end{enumerate}
For observation sequences which included two measurements of each ON position, the 
external errors were determined by using the differences of the three pairs of observations. 
After subtracting the ZL-background (modulated by detector drift effects), the two surface brightness measurements, $S_A$ and $S_B$, taken at each of the four positions POS$i$ (ON1, ON2, ON3 and the final pair of R2 measurements) were used to calculate the standard error of one measurement:
\[\sigma_{EXT} = \sqrt{\sum_{{\rm POS}i, i=1}^4 (S_{A} - S_{B})^2 / 8} \]

These values were then reduced by a factor of $1 / \sqrt{2}$ to obtain the standard error of the mean of the two measurements.\\

For the remaining bands, only one measurement was available for each ON
position; external standard errors were therefore estimated in all but two of these cases from the scatter of the three measurements of the reference positions R1 and R2. These error estimates also include any detector drift effects and intrinsic differences between R1 and R2, and are therefore upper limits. Drift effects dominated the L98 7.3- and 7.7-$\mu$m band measurements, and so for these final two cases, external errors were instead estimated from the scatter of the three ON measurements. The resulting external error estimates, expressed in MJy sr$^{-1}$, are listed in column 10 of Table 1.\\
\\
{\it Systematic (calibration) errors}\\
\\
At wavelengths $3.3 - 25$-$\mu$m where the calibration was based on the Zodiacal Light (Sect. 2.2), the relative filter-to-filter errors depend on the accuracy of the {\it shape} of the zodiacal emission spectrum and the statistical accuracy of the G~300.2~$-$~16.8 reference position measurements used in the calibrations (estimated to be $\sim$ 10 per cent). The absolute calibration accuracy also depends on the absolute zodiacal emission brightness in the {\it COBE}/DIRBE data. The adopted combined errors for the average ON {\it minus} REF cloud signal listed in the last column (14) of Table 1 were obtained by arithmetically adding the statistical external errors (column 11) and the filter-to-filter systematic (calibration) errors (column 12).

When estimating the filter-to-filter accuracies at $\lambda \ge 60 \mu$m and the absolute accuracies at
all wavelengths, the results of the \citet{klaas02} investigation of ISOPHOT accuracies were adopted. The estimated absolute accuracies are given in column 12 of Table 1. We note here that although the absolute accuracies for $\lambda \ge 60 \mu$m are $\sim$ 20 $-$ 25 per cent, the filter-to-filter uncertainties for a {\it single detector} (e.g. C2) may be substantially smaller due to the elimination of most of the sources of error ( see \citealt{delBurgo}, Sect. 4). This issue is important in constraining the FIR temperature fits (see Sect. 5.1 and Table 3). Unlike the shorter-wavelength data, the absence of any independent calibration curve to aid cross-calibration for the $\lambda \ge 60 \mu$m data dictated that the adopted combined errors for the average ON {\it minus} REF cloud signal listed in the last column (14) of Table 1 be obtained by arithmetically adding the statistical external errors (column 11) and the absolute errors (column 12).
   \begin{figure}
   \center
   \vspace{2mm} 
   \includegraphics[angle=0,width=8.5cm]{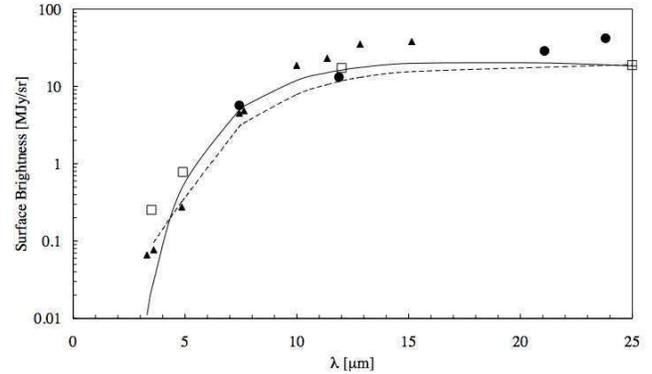}
      \caption{
Calibration of ISOPHOT photometry with COBE/DIRBE. The open squares represent the average of the COBE/DIRBE surface brightnesses at both REF positions for both observing dates. The filled circles represent the new photometry data after standard FCS1 calibration only. The solid curve represents the PHT-S and COBE/DIRBE-based averaged ZL spectrum adopted in this work. The dashed curve represents the ZL spectrum from L98. The small filled triangles represent the L98 photometry data after their standard FCS1 calibration only. The better agreement of the new data and the new ZL spectrum reflects improvements in both the normal FCS1 calibration and the techniques used in establishing the ZL spectrum.
              }
         \label{Fig2}
   \end{figure}
   \begin{figure*}
   \center
   \includegraphics[angle=0,width=16cm]{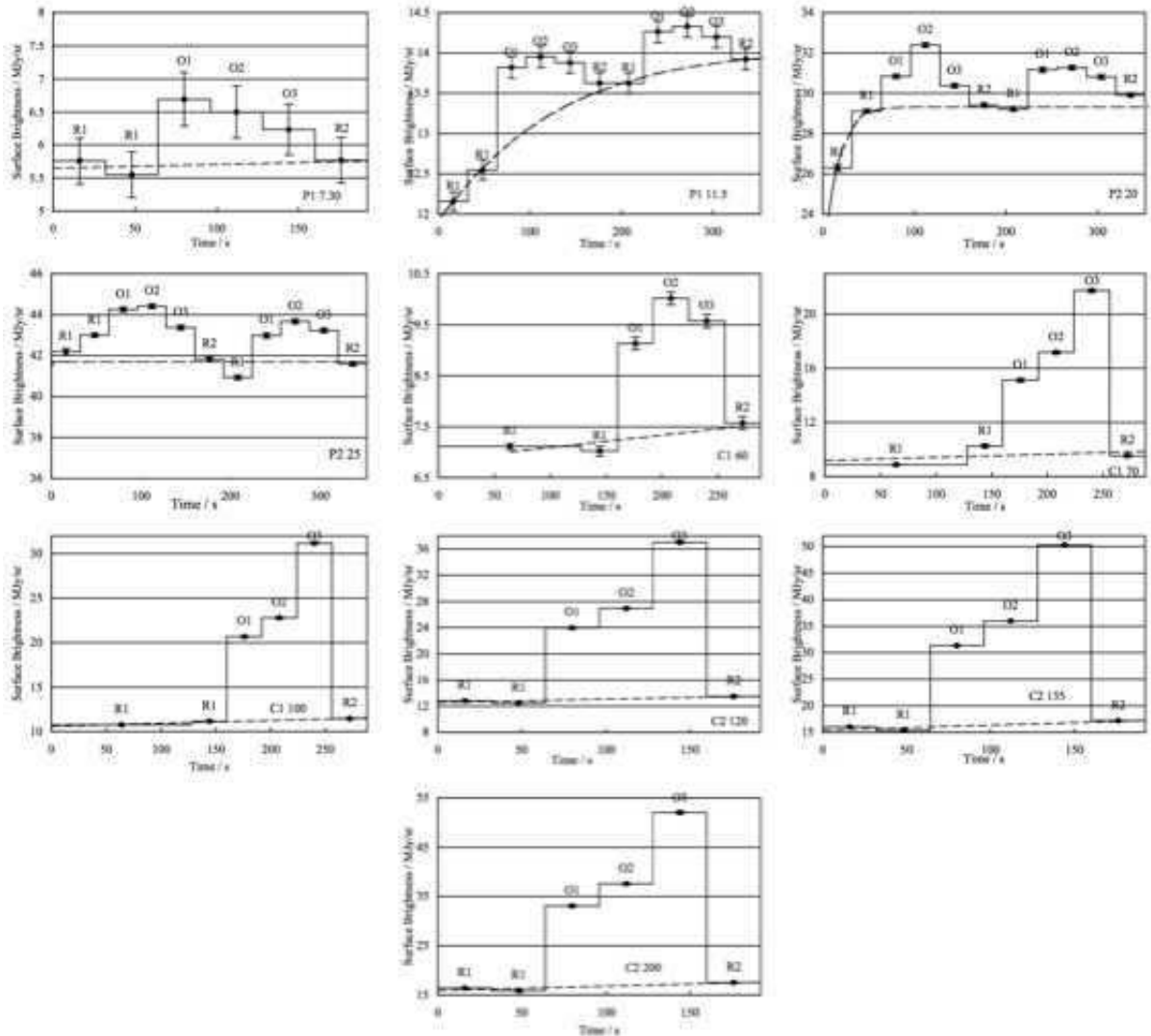}
      \caption{
Observed surface brightness in the different ISOPHOT filters prior to ZL recalibration and colour correction. The on-target and reference positions are designated as O1, O2, O3 and R1, R2 respectively. The measured values are plotted in the same sequence as they were measured. The dashed lines show fitted background zero levels. These were generally linear, except for the 11.5-, 20- and  25-$\mu$m filters, which used a fitted curve representing the detector drift. The surface brightness is given in the physical units (MJy/sr). The error bars shown indicate the internal statistical errors as obtained from the standard PIA reductions only. The time constants of the detector drift curves depend on the signal level, being smaller in the FIR detectors and when the signal is larger.
              }
         \label{Fig3} 
   \end{figure*}
\section{ISOPHOT results}

The observed in-band power of the cirrus emission at the three ON positions is given
in Table 2, combining the recalibrated data of L98 with the new data. Within errors, there is no disagreement between the pairs of P1\_7.3 measurements, despite the difference in aperture sizes.

Figs. 4 and 5 show the photometric data for the three observed positions as spectral energy
distributions. Qualitative differences in the relative strengths of the UIBs to the mid-
and far-IR grain emission are immediately apparent. The spectrum at position ON1
clearly exhibits relatively strong UIBs, with the 7.7- and 11.3-$\mu$m bands
demonstrating the importance of their carriers at this position. The UIBs are clearly
the dominant factor that gives rise to a blue 12 $\mu$m/100 $\mu$m {\it IRAS} colour at
ON1. At the ON2 position, the strongest of the UIBs (at 7.7 $\mu$m) and the thermal
far-IR emission peak are at approximately the same level. Although still weaker than
either of these, the mid-IR fluxes between 20 and 60 $\mu$m are at their relative
maxima here, again in accordance with the high 25-$\mu$m {\it IRAS} value. The
spectrum of position ON3 is dominated by the far-IR thermal emission from the larger
dust grains, and the UIBs are comparatively weak. Another notable feature of the
data is the clear presence of some form of underlying continuum emission,
centred on a position longward of $\sim$10 $\mu$m.

   \begin{figure*}
   \center
   \includegraphics[angle=0,width=11.2cm]{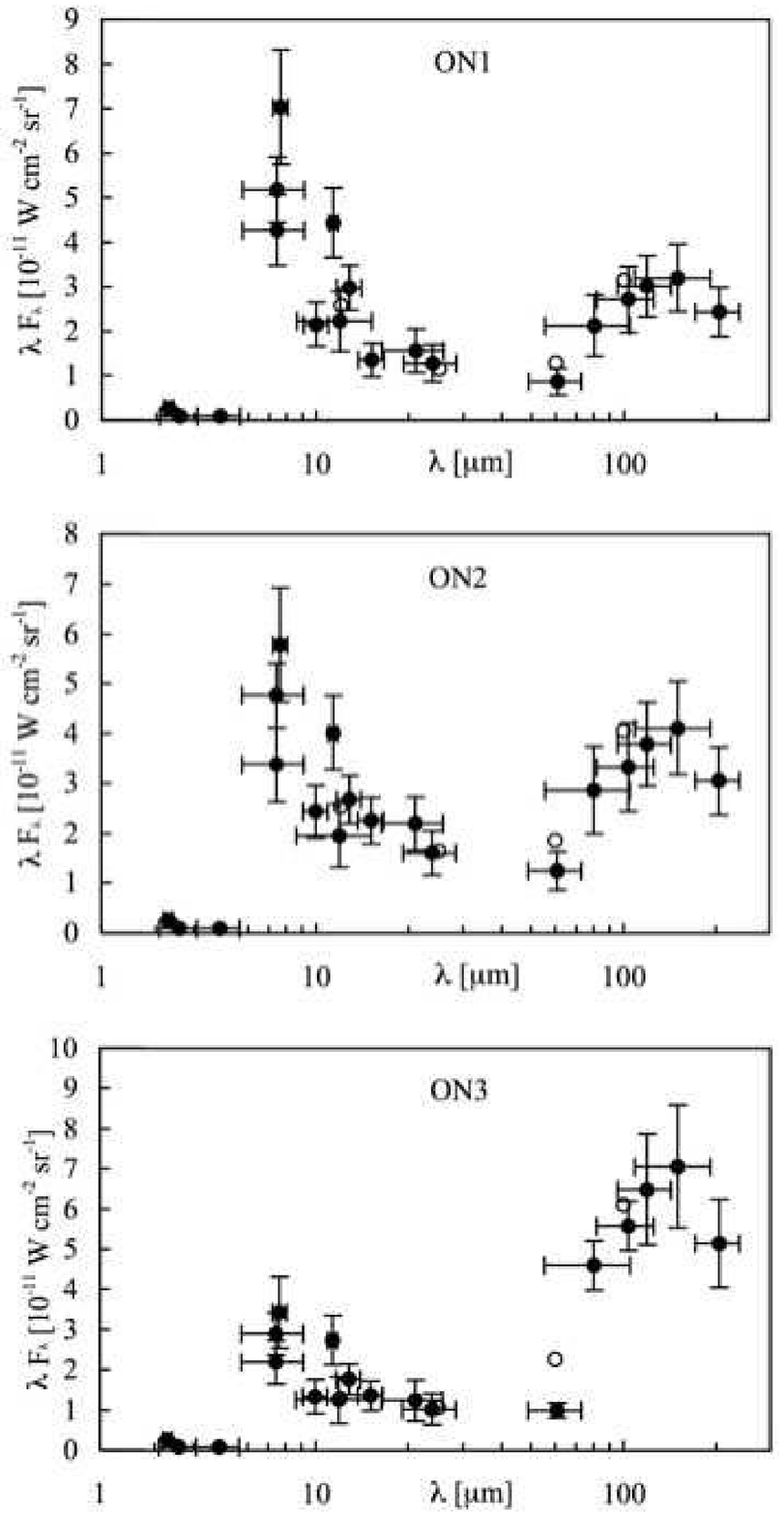}
      \caption{
Background-subtracted ISOPHOT photometry of the ON1, ON2 and ON3 positions (filled circles). The four IRAS fluxes are included for comparison (open circles).
ISOPHOT data at $\lambda \ge 20 \mu$m have been colour corrected. The horizontal bars reflect the effective filter band widths.
              }
         \label{Fig4}
   \end{figure*}
   \begin{figure*}
   \center
   \includegraphics[angle=0,width=15cm]{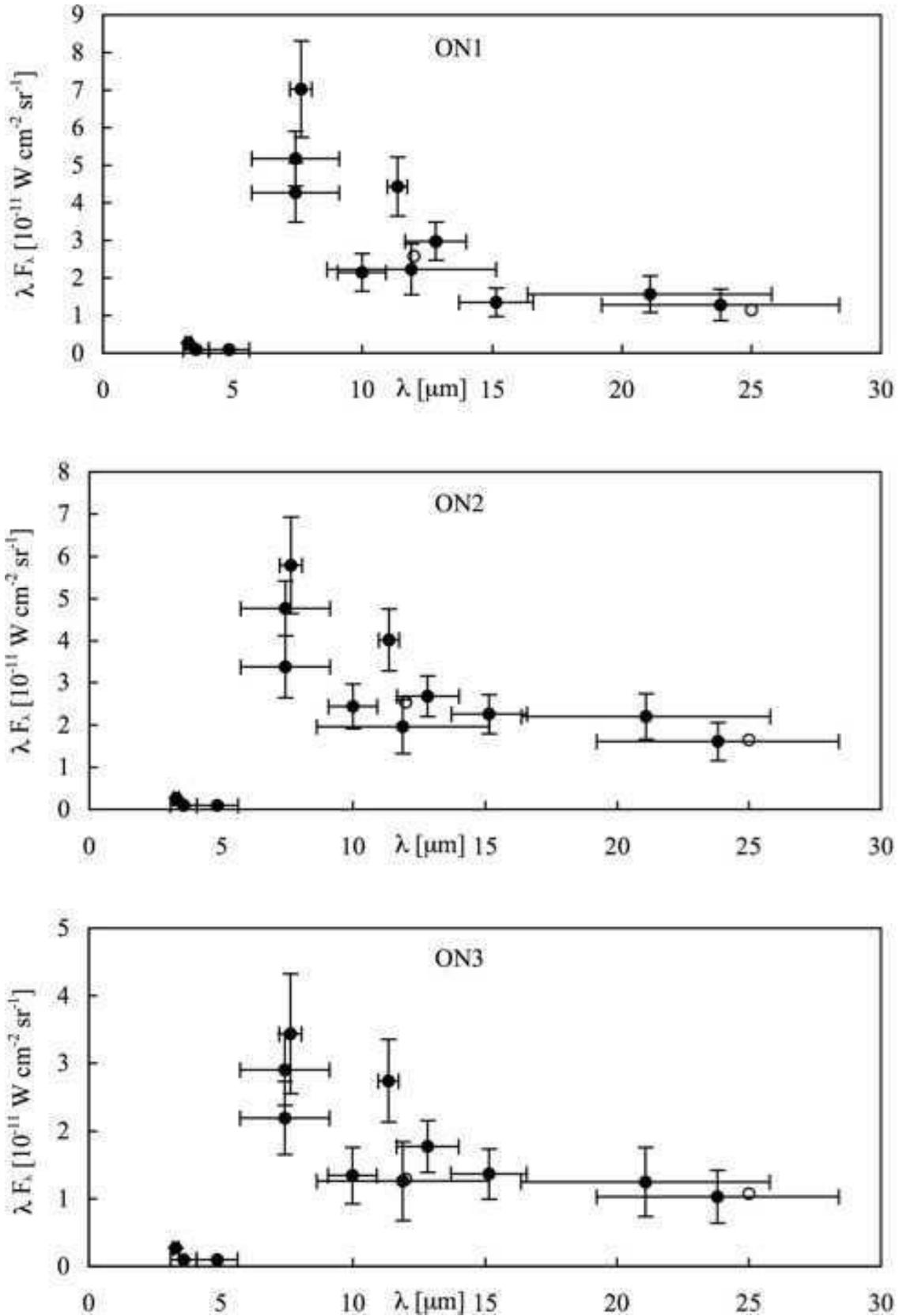}
      \caption{
Background-subtracted ISOPHOT photometry of the UIB wavelength range at the ON1, ON2 and ON3 positions (filled circles). The two IRAS fluxes are included for comparison (open circles). ISOPHOT data at $\lambda \ge 20 \mu$m have been colour corrected. The horizontal bars reflect the effective filter band widths.
              }
         \label{Fig5}
   \end{figure*}

\begin{table}
\center
   \caption[2]{Positions, {\it IRAS} filter band ratios, extinctions and in-band emission powers of the
cirrus cloud (bands+continuum) as observed with ISOPHOT in the different filter bands. Visual extinctions
quoted here are based on the 2MASS mapping described in Sect. 4. The in-band powers are obtained
here by multiplying the monochromatic fluxes by the filter widths, and are relative to a zero
level determined at the reference positions (REF1: $\alpha = 11^{\rm{h}} 46^{\rm{m}} 18.9^{\rm{s}},
\delta=-78^{\circ} 46' 33.2''$, REF2: $\alpha = 12^{\rm{h}} 06^{\rm{m}} 23.4^{\rm{s}},
\delta=-79^{\circ} 22' 07.0''; J2000.0$). The error estimates given in
parentheses include (external) statistical errors
and a relative filter-to-filter error or absolute accuracy (see column 14 of Table 1).
}
      \label{Table2}

\begin{tabular}{llll}
\hline
  Position &        ON1 &        ON2 &        ON3 \\
\hline
\multicolumn{ 1}{l}{$\alpha , \delta$ (J2000.0)} & $11^{\rm{h}} 52^{\rm{m}} 08.3^{\rm{s}}$ & $11^{\rm{h}} 48^{\rm{m}} 24.4^{\rm{s}}$ & $11^{\rm{h}} 55^{\rm{m}} 33.8^{\rm{s}}$ \\

\multicolumn{ 1}{l}{} & $-79^{\circ} 09' 32.5''$ & $-79^{\circ} 17' 59.7''$ & $-79^{\circ} 20' 54.0''$ \\
\hline
   $A(V)$ & $1.8 \pm 0.2$ & $1.9 \pm 0.2$ & $2.9 \pm 0.2$ \\
   $A(V)$ (ON - REF)$^{\mathrm{a}}$ & $1.2 \pm 0.2$ & $1.3 \pm 0.2$ & $2.3 \pm 0.2$ \\
\hline
$I_{\nu} (12) / I_{\nu} (100)$ &       0.11 &       0.08 &      0.029 \\

$I_{\nu} (25) / I_{\nu} (100)$ &      0.094 &      0.105 &      0.044 \\

$I_{\nu} (60) / I_{\nu} (100)$ &      0.244 &      0.273 &      0.223 \\

$I_{\nu} (12) / I_{\nu} (25)$ &       1.17 &      0.762 &      0.659 \\
\hline
Filter Band & \multicolumn{ 3}{l}{In-band power, $P$ [$10^{-12}$ W cm$^{-2}$ sr$^{-1}$]} \\
\hline
   P(3.29)$^{\mathrm{b}}$ & $\le0.18$ & $\le0.18$ & $\le0.18$ \\

    P(3.6)$^{\mathrm{b}}$ & $\le0.27$ & $\le0.27$ & $\le0.27$ \\

   P(4.85)$^{\mathrm{b}}$ & $\le0.31$ & $\le0.31$ & $\le0.31$ \\

P(7.3)$^{\mathrm{c}}$ &  23.5(4.0) &  21.7(4.1) &  13.2(3.2) \\

P(7.3)$^{\mathrm{d}}$ &  19.4(3.0) &  15.4(2.4) &  10.0(1.9) \\

    P(7.7) &  7.7(1.4) &  6.4(1.3) &  3.8(1.0) \\

     P(10) &  4.0(0.9) &  4.5(1.0) &  2.5(0.8) \\

   P(11.3) &  3.0(0.5) &  2.7(0.5) &  1.9(0.4) \\

   P(11.5) &  12.2(3.7) &  10.7(3.5) &   6.9(3.2) \\

   P(12.8) &  5.4(0.9) &  4.9(0.9) &  3.2(0.7) \\

     P(16) &  2.6(0.7) &  4.3(0.9) &  2.6(0.7) \\

     P(20) &  7.0(2.2) &  9.8(2.4) &  5.6(2.3) \\

     P(25) &  4.9(1.6) &  6.2(1.7) &  4.0(1.5) \\

     C(60) &  3.4(1.2) &  4.9(1.5) &  3.9(0.7) \\

    C(70) & 13.1(4.2) & 17.7(5.4) &  28.4(3.8) \\

    C(100) & 11.4(3.1) & 14.0(3.8) &  23.6(2.6) \\

    C(120) & 12.0(2.7) & 15.1(3.3) &  25.8(5.5) \\

    C(135) & 17.6(4.1) & 22.6(5.1) & 38.9(8.4) \\

    C(200) &  8.0(1.8) & 10.0(2.2) &  16.9(3.6) \\
\hline
\end{tabular}
\begin{list}{}{}
\item[$^{\mathrm{a}}$] $A(V)$ at reference position 1 = $0.7 \pm 0.1$; $A(V)$ at reference position 2 = $0.5 \pm 0.1$.
\item[$^{\mathrm{b}}$] Upper limits for bands+continuum, taken from table 3 of L98.
\item[$^{\mathrm{c}}$] L98; 180'' aperture.
\item[$^{\mathrm{d}}$] This work; 99'' aperture.
\end{list}
\end{table}
\section{Optical extinction and scattered light}

When describing the ISM in G~300.2~$-$~16.8, knowledge of the $A(V)$ and $R(V)$
values are needed, where $R(V)\equiv A(V)/E(B-V)$.
Star count data, i.e. the number of stars per square degree $N(m)$ brighter than magnitude $m$,
in the $B_J$- and $I$-bands have been extracted from the
USNO-B1.0 archive \citep{monet}. We have used data from the second epoch
SERC-$J$ ($B$-band) and SERC-$I$ ($I$-band) surveys. The value of the extinction $A$ is
derived with the formula \\
\\
$A = (1/k)$ log ($D_{ref} / D)$ \\
\\
where $k$ is the slope of the log $N(m)$ vs. $m$ relationship, $D_{ref}$ and $D$ are the stellar
number densities at the reference area and at the cloud area. The values of $k$ and $D_{ref}$ have
been determined from the reference area.

\begin{figure}
   \center
   \includegraphics[bb=5 5 590 414,angle=0,width=9cm]{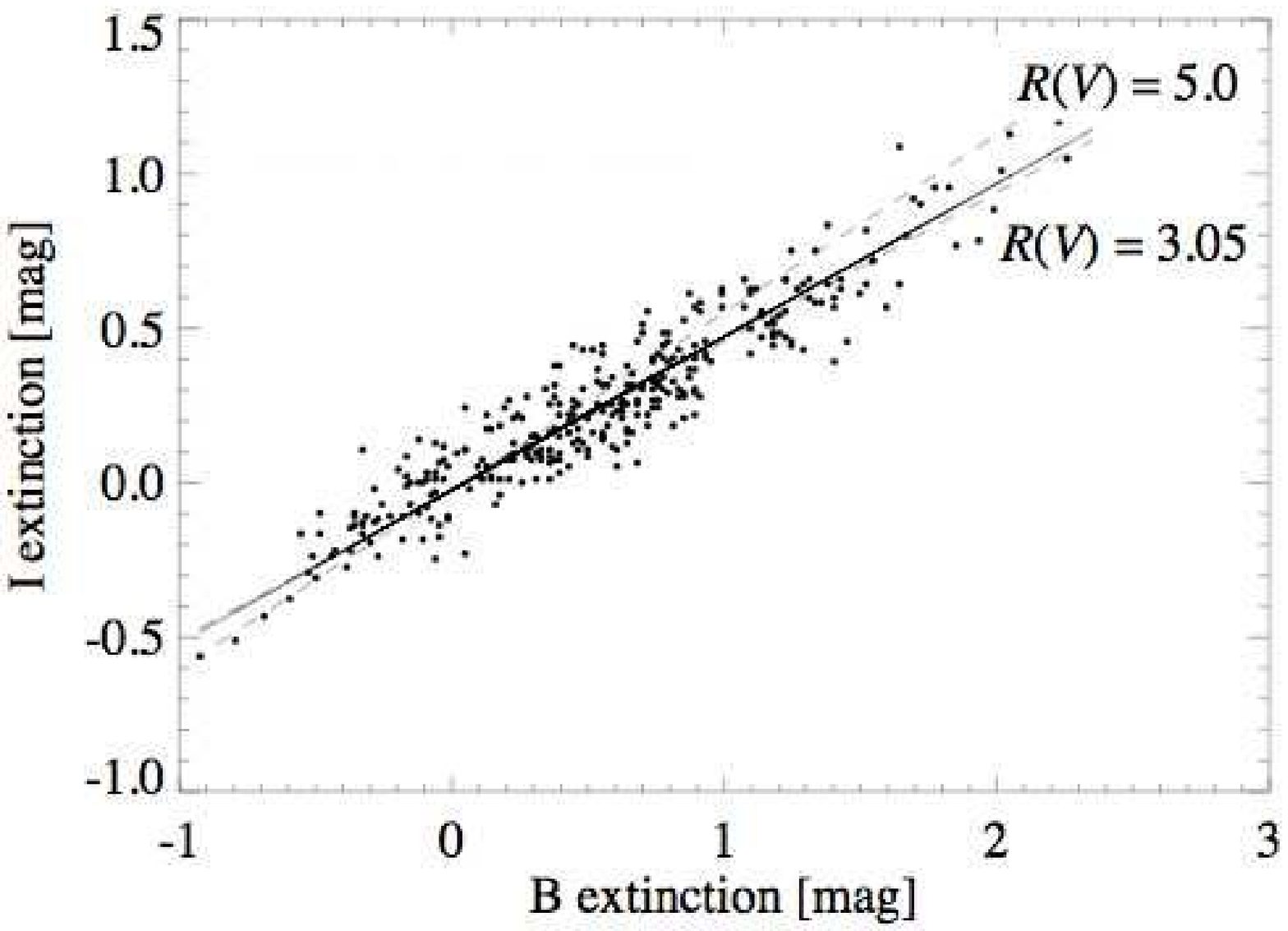}
      \caption{Extinctions obtained from star counts using the $B_{J}$ and $I$ band data of the USNO-B1.0 archive \citep{monet}. The slopes for $R(V)$ = 3.05 and $R(V)$ = 5.0 are shown as dashed lines for comparison. The slope of the fitted line (continuous line) suggests $R(V)\sim 3.1$ for G~300.2~$-$~16.8.}
         \label{Fig6}
   \end{figure}

We have used $4700 \AA $ and $8100 \AA $ for the effective wavelengths of the $B_J$
and $I$ bands respectively \citep{gullixson}.
The observed relation between $B_J$ - and $I$-band
extinctions is shown in Fig. 6. This yields the relation
$A(I) = 0.49 \pm 0.01 A(B_{J})$.
Using the \citet{cardelli} extinction curve for the case
of $R(V) = 3.05$ this should be
$A(I) = 0.48 A(B_{J})$
whereas for the $R(V) = 5.0$ case, we would expect
$A(I) = 0.58 A(B_{J})$. \\
The measured slope therefore suggests that $R(V) \sim 3.1$ for these sightlines, as would
be expected for a diffuse cloud. It should be noted, however, that the photometric
calibration of the USNO-B1.0 catalogue is still at a preliminary stage. This may give
rise to an uncertainty in the $A(I)/A(B_{J})$ slope which could potentially dominate the
inherent statistical errors as a consequence of calibration variations between survey
plates. Consequently, only fields from one plate were included in the analysis.

The 2MASS $JHK_{S}$ data can be used to derive near-IR colour excesses of stars
visible through G~300.2~$-$~16.8, and hence derive NIR extinction. We have applied the
optimized multi-band technique of \citet{lombardi}. Stellar
density for the stars which are detected at all three bands is constant throughout the
map, with an average of 2.4 stars arcmin$^{-2}$.

The intrinsic colours of stars in the reference areas have the mean and standard
deviations $(J - H)_{0} = 0.45 \pm 0.19$ and $(H - K_{S})_{0} = 0.12 \pm 0.21$. For
the ratio of visual extinction to colour excess, we have used $A(V) / E(J - H) = 9.90$
and $A(V) / E(H - K_S) = 14.70$, which correspond to $R(V) = 3.05$.

   \begin{figure*}
   \center
   \vspace{20mm}
   \includegraphics[angle=0,width=16cm]{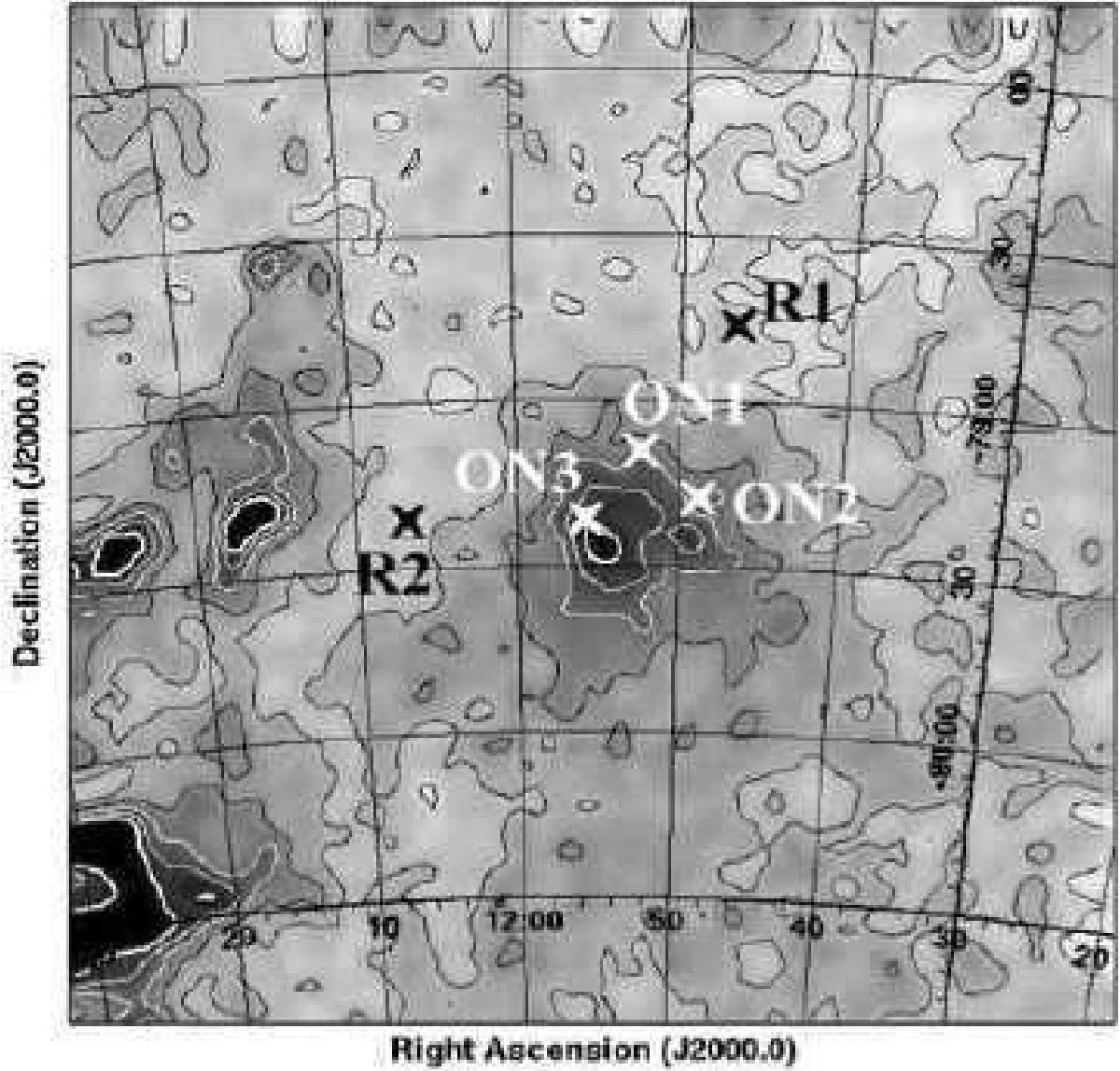}
      \caption{
$A(V)$ map obtained from the 2MASS $JHK_{S}$ colour excesses of stars visible through G~300.2~$-$~16.8 via the multi-band colour excess method. The pixel size is 1.5 arcmin with a Gaussian FWHM = 4.0 arcmin used as a weighting function for the individual extinction values. The contours are from $A(V)$ = 0.5 to 3.5 mag in steps of 0.5 mag, with the contour or the local maximum near the centre of the field (near ON3 in G~300.2~$-$~16.8) at the 3.0 level.
              }
         \label{Fig7}
   \end{figure*}

The pixel size of the extinction map was chosen to be 1.5 arcmin. The value of the extinction
in each map pixel was derived from the individual extinction values of stars by
applying the sigma-clipping smoothing technique of \citet{lombardi}, and
using a Gaussian with FWHM = 4.0 arcmin as a weighting function for the individual
extinction values. There are two sources of variance in the extinction map; variance of
the intrinsic colours $(J - H)_{0}$ and $(H - K)_{0}$, and variance of the observed
magnitudes of the field stars. In the $A(V)$ map, the former source dominates with values of
$\sim$0.15 magnitudes 1$\sigma$ error per pixel, while the latter source typically
gives $\sim$0.05 magnitudes 1$\sigma$ error per pixel. The visual extinction map is shown in
Fig. 7. This produced extinction values at the three ON positions of: \\
\\
$A(V)$ (ON1) $= 1.8 \pm 0.2$ \\
\\
$A(V)$ (ON2) $= 1.9 \pm 0.2$ \\
\\
$A(V)$ (ON3) $= 2.9 \pm 0.2$ \\
\\
In order to constrain the energy budget of G~300.2~$-$~16.8 (see Sect. 6
and 8.7), we have conducted optical $B_{J}$- and
$R$-band surface photometry of the cloud. G~300.2~$-$~16.8 appears
on the same ESO/SERC sky survey plates as the Thumbprint Nebula (TPN). The
$R$- and $B$-band photometry data presented in \citet{lehtinen98} and \citet{fitzgerald}
for the TPN were used to convert the photographic density values for G~300.2~$-$~16.8 into
$R$- and $B$-band surface brightnesses.

In addition, the $UBVRJHK$ surface photometry data for the TPN in
\citet{lehtinen98} were taken as a template with which to estimate the $UBVRJHK$ SEDs for the
G~300.2~$-$~16.8 positions (see Fig. 10). The different optical depths, $\tau_{\lambda}$, for the TPN
and G~300.2~$-$~16.8 positions were taken into account by applying scaling factors of
$(1-e^{-\tau_\lambda })$, and scaled to provide a best fit to the brightnesses at $B_{J}$ and $R$.
The $\tau_{\lambda}$ values for G~300.2~$-$~16.8 were determined from the
measured $A(V)$ values using $R(V) = 3.05$. From a comparison between the extinction-adjusted
TPN template and the photometric values for G~300.2~$-$~16.8, it can be concluded that:

\begin{enumerate}
\item The surface brightness SED template from the TPN
reproduces the $B_{J}$ and $R$ measurements of G~300.2~$-$~16.8 very well.
It therefore appears reasonable to also use the TPN-based template as a first approximation for
G~300.2~$-$~16.8 shortward of the $B_{J}$- and longward of the $R$-band.\\

\item Using this purely empirical approach, we are not assuming
that the dust scattering properties in G~300.2~$-$~16.8 and the TPN are the same.
There could also be a contribution by the Extended Red Emission (ERE)
which, because of the consistency of the $B_{J}$- and $R$-band values,
would have to be at similar relative levels in these two objects.\\
\end{enumerate}
Consequently, the resultant values were assumed to broadly represent the optical surface brightness of G~300.2~$-$~16.8.

\section{Spectral modelling}
\subsection{Semi-empirical UIR band and FIR dust emission models}
Based on the observed spectral energy distributions, modelling of the UIB emission is
possible, if spectral feature profiles as observed in other Galactic emission regions are
assumed. A number of types of line profile have previously been used by various
authors attempting to describe the UIB spectrum, including Gaussian and Drude
profiles (see table 7 of \citealt{li01b} for a summary).  \citet{boulanger98}
and \citet{mattila99} found that astronomical spectra of the UIR bands could be
well fitted to within ISOPHOT-S/ISOCAM-CVF resolutions using Cauchy/Lorentzian
distribution functions. The advantage of the use of these profiles in model fits over
Gaussians is that the need for an underlying plateau continuum emission between 5
and 9.5 $\mu$m is reduced, as the superposition of the wings of the Cauchy curves can
account for more of the plateau emission. The curves may also be in accordance with
the physical emission mechanisms of large molecules at high temperatures, with the
Cauchy profiles being associated with the short excitation lifetime of vibrational
states \citep{boulanger98}.

Our new data do not cover the 3.29- and 6.2-$\mu$m UIBs, and so the ISOPHOT
photometry for these particular bands has not been improved upon since L98. On the
other hand, an ISOCAM-CVF spectrum of a position coincident with the ON1 position has
since been presented by \citet{boulanger98}. This spectrum may therefore be taken as
indicative of the relative heights of the UIR bands, allowing reasonable assumptions
to be made about the relative heights of some of the UIBs. A combination of this
foreknowledge and our ISOPHOT photometry allows us to produce semi-empirical
models of the UIR band emission spectra for our three positions in G~300.2~$-$~16.8.

The 3.29-, 7.7-, 11.3- and 12.7-$\mu$m bands are covered by individual filters, which
are sufficiently narrow as to enable a good characterization of the band heights. In
contrast, the broad P7.3 filter spans a wavelength range encompassing the 6.2-, 7.7-
and 8.6-$\mu$m bands, but still provides some constraints on emission. The model
spectra are derived under the assumption of a Cauchy profile for each of the six
expected major UIR bands. In addition to these, a continuum
baseline function of the silicate dust emission spectrum from $2 - 18 \mu$m, obtained
from the physical modelling of the ON1 sightline (see Sect. 5.2 below), was also
incorporated. The derived UIR band strengths are not strongly dependent on this
specific choice of baseline function. Very similar results are obtained with other, more
ad hoc baseline functions, e.g. a modified blackbody with $T\sim$300K, peaking near
10 $\mu$m, so the UIB modelling results are not dependent on an assumed
enhancement of the small silicates population.  Together, these functions produced a
short wavelength model spectrum. This spectrum was then convolved with each of the
ISOPHOT filter response curves in turn, and the resultant flux levels compared with
the in-band power photometry values obtained with the P7.3, P7.7, P10, P11.3, P11.5,
P12.8, P16 and P20 filters (Table 2). Parameters of the curves were then iteratively
adjusted to minimize the $\chi^{2}$ statistic. Central wavelengths were determined
first by fitting under the assumption of a spectrum similar to that of
\citet{boulanger98}, and then fixed for subsequent detailed fitting. UIB widths for the
3.3-, 6.2-, 7.7-, 8.6- and 11.3-$\mu$m bands in the diffuse medium were taken from
\citet{kahanpaa}, and the width of the 12.7-$\mu$m band was allowed
to vary between 0.27 and 1.18 $\mu$m, in accordance with the broad range of
previously-reported function widths listed in \citet{li01b}. Due to the lack of
narrow-band coverage of the 6.2-$\mu$m band by our dataset, an independent
measurement of this band was not possible, and so height ratio constraints were
applied, which were determined by comparing the Boulanger ISOCAM-CVF measurements
taken near ON1 with average values obtained from \citet{boulanger98} for
Ophiucus and NGC 7023 and \citet{mattila99} ISOPHOT-S observations of NGC
891. The two UIB height ratio constraints applied specified that 0.76 $\le$ (6.2/7.7)
$\le$ 0.77 and 0.20 $\le$ (8.6/7.7) $\le$ 0.29.

The three longest-wavelength photometry points, C120, C135 and C200 are fitted
using a modified blackbody function of the form $\nu^{2} B(\nu)$. This function is
individually convolved with each of the three ISOPHOT filter response curves in turn,
and the temperature and scaling are adjusted to best fit the photometry. In all three
cases, a temperature of $\sim$17.5 K was obtained (Table 3). These temperatures
reflect the average properties of the cloud, and do not preclude the possibility of
colder grains near the cloud centre.

   \begin{figure*}
   \center
   \includegraphics[angle=0,width=14cm]{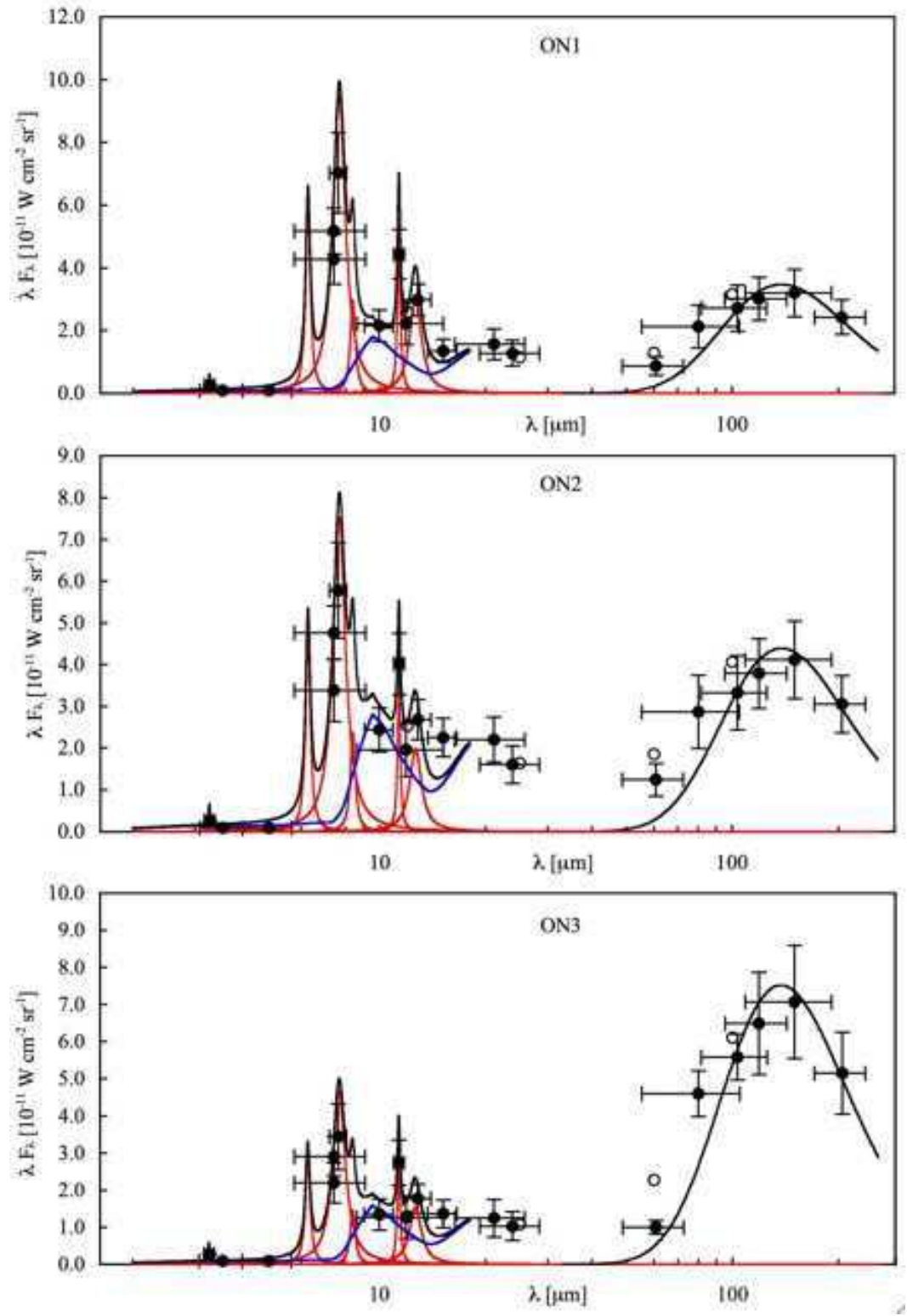}
      \caption{
A modified blackbody fit of the form  $\nu ^{2} B_{\nu} (T)$ to the $120-200$ $\mu$m data points is shown as a black line spanning the $40-250$ $\mu$m wavelength range. Cauchy band profiles fitted to the $6.2-12.7$ $\mu$m UIBs are shown in red. A \citet{li01b}-style silicate continuum is shown in blue for the $6.2-20$ $\mu$m range, and a combination of these is shown as a black line for $3-20$ $\mu$m.
              }
         \label{Fig8}
   \end{figure*}

The resulting fits are shown in Fig. 8 and the parameters are listed in Table 3. It can
be clearly seen from Fig. 8 that in all three cases, the middle range of the spectrum
($\sim$25 $-$ 70 $\mu$m) exhibits additional emission that cannot be accounted for by these
two populations alone. Consequently, a third grain component in the infrared
is required. As this arises due to non-equilibrium transient heating processes, detailed
grain population modelling is required, as is described in Sect. 5.2. We also note
that although the continuum near 10 $\mu$m does not appear to correlate strongly with
the UIBs, Fig. 8 may indicate a correlation between the 10-$\mu$m continuum and the mid-IR
emission from $\sim$16 $-$ 25 $\mu$m.

\begin{table} 
\center
   \caption[3]{Results of semi-empirical model fitting. Overall
errors on the peak heights are estimated to be $\sim 20$ per cent, with relative band-to-band
errors $\sim 10-15$ per cent. $F(\lambda )$ values are quoted in units of $10^{-12}$W cm$^{-2} \mu \rm{m}^{-1} \rm{sr}^{-1}$.}
      \label{Table3}
\begin{tabular}{lllll}
\hline 
Feature & Fit parameter &        ON1 &        ON2 &        ON3 \\
\hline
3.3-$\mu\rm{m}$ & Central $\lambda$ $[\mu\rm{m}]$ &       3.29 &       3.29 &       3.29 \\

band$^{\mathrm{a}}$ & Width$[\mu\rm{m}]$ &       0.04 &       0.04 &       0.04 \\

           & $F(\lambda)$ Height &     $\le 1.47$ &     $\le 1.47$ &     $\le 1.47$ \\
\\
6.2-$\mu\rm{m}$ & Central $\lambda$ $[\mu\rm{m}]$ &       6.27 &       6.27 &       6.27 \\

band       & Width$[\mu\rm{m}]$ &       0.24 &       0.24 &       0.24 \\

           & $F(\lambda)$ Height &       9.5 &       7.6 &       4.7 \\
\\
7.7-$\mu\rm{m}$ & Central $\lambda$ $[\mu\rm{m}]$ &       7.68 &       7.68 &       7.68 \\

band       & Width$[\mu\rm{m}]$ &       0.76 &       0.76 &       0.76 \\

           & $F(\lambda)$ Height &      12.3 &        9.8 &       6.0 \\

\\
8.6-$\mu\rm{m}$ & Central $\lambda$ $[\mu\rm{m}]$ &        8.40 &        8.40 &        8.40 \\

band       & Width$[\mu\rm{m}]$ &       0.33 &       0.33 &       0.33 \\

           & $F(\lambda)$ Height &       3.5 &       2.8 &       1.7 \\
\\
11.3-$\mu\rm{m}$ & Central $\lambda$ $[\mu\rm{m}]$ &      11.35 &      11.35 &      11.35 \\

band       & Width$[\mu\rm{m}]$ &       0.27 &       0.27 &       0.27 \\

           & $F(\lambda)$ Height &       4.6 &       2.9 &       2.3 \\
\\
12.7-$\mu\rm{m}$ & Central $\lambda$ $[\mu\rm{m}]$ &      12.64 &      12.64 &      12.64 \\

band       & Width$[\mu\rm{m}]$ &       1.18 &       1.18 &       1.18 \\

           & $F(\lambda)$ Height &       2.4 &       1.6 &       1.2 \\
\hline
7.7/11.3-$\mu\rm{m}$ & &       $2.7 \pm 0.4$ &      $3.4 \pm 0.5$ &        $2.6 \pm 0.4$ \\
band ratio & & & & \\
\hline
\multicolumn{ 2}{l}{10-$\mu\rm{m}$ continuum level} & $2.3 \pm 0.4$ & $3.1 \pm 0.6$ & $1.8 \pm 0.4$ \\
\multicolumn{ 3}{l}{(average of $F(\lambda)$ over $9.5-10.5$ $\mu\rm{m}$ range)} & & \\
\hline
\multicolumn{ 2}{l}{Classical grains: equilibrium} &       17.4 &       17.4 &       17.5 \\

\multicolumn{ 2}{l}{temperature, $T$ [K]} & $\pm^{1.8}_{2.2}$ & $\pm^{1.7}_{1.4}$ & $\pm^{1.6}_{1.3}$ \\
\hline
\end{tabular}
\begin{list}{}{}
\item[$^{\mathrm{a}}$] Adopted from L98
\end{list}
\end{table}
\subsection{Physical modelling}
An alternative approach to the semi-empirical fit described in Sect. 5.1 is to
construct a quantitative physical model of the cloud. Although a full numerical
analysis is beyond the scope of this paper, we present here a first approximation
of a three-component numerical model for G~300.2~$-$~16.8. A detailed numerical
model will be presented in a subsequent paper.

The numerical modelling code based on the Monte Carlo method
(see \citealt{juvela03} for full details) was applied. Spherical symmetry
was assumed and clumping effects were neglected. The numerical model was applied to
a grid of spherically-symmetric concentric shells. Due to the simple nature of the
model, the library method of \citet{juvela03} was not used.

The Solar neighbourhood interstellar radiation field of \citep{mathis83} as modified by
\citet{lehtinen98} was adopted. The model clouds were discretized into 50
shells. The density was constant within the innermost 10 per cent of the cloud
radius and in the outer parts the density decreased following power
law $n\sim r^{-1.0}$. For each one of the three positions, ON1, ON2 and ON3, a separate cloud model was adopted, centred on each position in turn.

The dust properties were based on the \citet{li01b} model. The dust consists of
silicate grains (sizes $a > 3.5$ \AA), graphite grains ($a>50$\AA ) and PAHs (from $a=
3.5 \AA$ to $\sim50 \AA$) with optical properties corresponding to warm ionized medium.
We have treated larger graphite grains ($a > 50$ \AA ) and smaller graphite grains
including the PAHs (from $a = 3.5 \AA$ to $\sim50 \AA$) as separate populations. As in Li
\& Draine, below 50 \AA ~there is a smooth transition from the optical properties of
graphite grains to PAH properties. The Li \& Draine model was modified by removing
emission features around 20 $\mu$m due to the lack of any such observed features in
{\it ISO} SWS spectra. Between 15 $\mu$m and 36 $\mu$m, the absorption and scattering
cross sections were replaced with values obtained with linear interpolation.

Monte Carlo methods were used to simulate the radiation field in the cloud.
Temperature distributions of the grains were calculated in each cell based on the
simulated intensity of the field (see \citealt{juvela03}). The dust emission spectrum,
consisting of some 300 frequency points, was calculated towards the centre of the
model cloud, and finally convolved with ISOPHOT
filter profiles for direct comparison with observations.

The grain size distributions were modified in order to find the best correspondence
between the model predictions and the observations. The size distributions of silicate
grains and larger graphite grains were modified by a factor $\propto a^{\Delta
\gamma}$, i.e. by changing the exponent of the size distribution by $\Delta \gamma$.
The value of $\Delta \gamma$ was limited to a range [-1.5, +1.5]. PAHs and small
graphite grains below 50\AA  were treated as a separate population for which the
shape of the size distribution was not modified. The total abundance of each of the
three grain populations was treated as a free parameter.

   \begin{figure*}
   \center
   \includegraphics[angle=0,width=11.5cm]{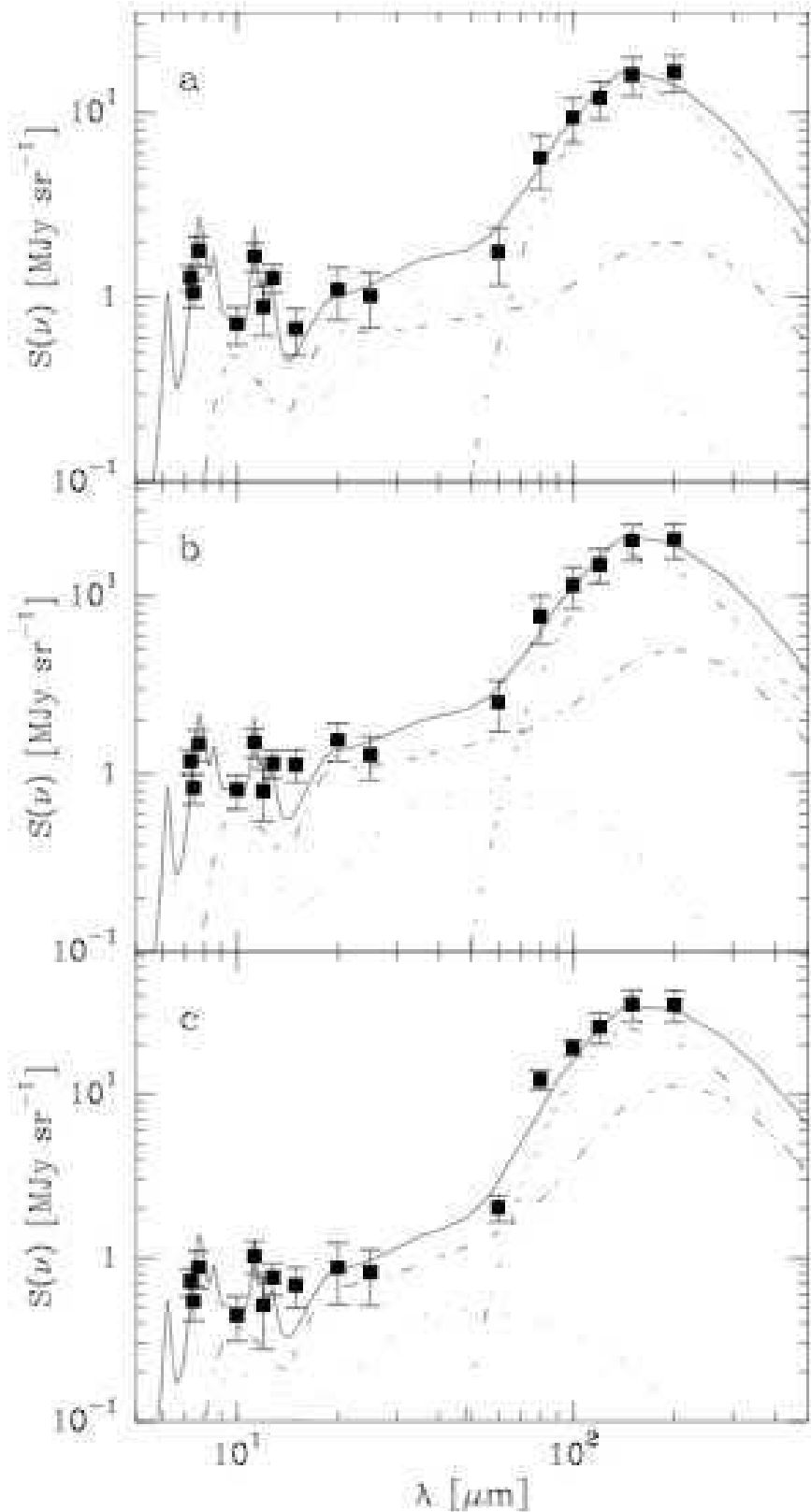}
      \caption{
Results of the physical modelling. The solid line shows the model spectra computed using the parameters listed in Table 4. The filled squares with the error bars are the observations and the open circles are the model predictions obtained by convolving the model spectrum with the corresponding filter profiles. The remaining curves show the components of the model spectrum: silicates (dashed line), large graphite grains (dash-dotted line) and small graphite grains including PAHs (dotted line).
              }
         \label{Fig9}
   \end{figure*}

Figure 9 shows the fits, and Table 4 lists the obtained model parameters. The models
preferred distributions with large graphite grains with $\Delta \gamma$ close to the
allowed maximum value of 1.5 in all cases. The average size of the silicate grains was
correspondingly decreased. The abundance of PAHs shows a strong variation, ranging from an
overabundance by a factor of $\sim 3$ at the `halo' position ON1 to an underabundance by a factor of $\sim 2$ at the centre position ON3. The derived visual extinction values giving the best fit
agree with the $A(V)$ (ON - REF) values derived in Sect. 4 to within $\leq 26$ per cent.

\begin{table}
\center
   \caption[4]{Parameters resulting from the physical modelling of the three source positions. The grain size distributions relative to the \citet{li01b}
model were modified by scaling the abundance by a factor $X$ and by changing the
exponent of the size distribution, $\propto (a)^{\Delta \gamma}$. The allowed
range for the values of $\Delta \gamma$ was [$-$1.5, 1.5]. For PAHs, only the
abundance was changed. The abundances are given relative to the silicate
component. The total dust column density resulting from the model fit is given
as $A(V)$ ($R(V) = 3.05$ has been used). The final row gives the total
silicate abundance requirement, i.e. a factor by which the \citet{li01b}
model abundances have to multiplied.
}
      \label{Table4}
\begin{tabular}{lllll}
\hline
           &            &        ON1 &        ON2 &        ON3 \\
\hline
       PAH &       $X$ &       2.84 &       1.13 &       0.54 \\
\\
  Graphite &       $X$ &       0.57 &       0.31 &       0.73 \\

           & $\Delta\gamma$ &       1.47 &       1.48 &       1.50 \\
\\
  Silicate &       $X$ &       1.00 &       1.00 &       1.00 \\

           & $\Delta\gamma$ &    $-1.50$ &    $-1.38$ &    $-1.00$ \\
\hline
$A(V)^{\mathrm{a}}$ [mag] &            &       0.89 &       1.39 &       1.76 \\
\hline
Silicate Abundance &            &       1.50 &       3.25 &       2.09 \\
\hline
\end{tabular}
\begin{list}{}{}
\item[$^{\mathrm{a}}$] These values correspond to the $A(V)$ (ON - REF) values listed in Table 2.
\end{list}
\end{table}

In our modified Li \& Draine-type model, the PAHs cannot explain the 10-$\mu$m
continuum level as it stands. Excluding the unlikely possibility of another strong
emission feature longward of 20 $\mu$m, either extra carbon
absorption or small silicates must be invoked, in the following ways:
\begin{enumerate}
\item The 10-$\mu$m carbon absorption coefficient is increased, which effectively
smoothes the continuum near 16 $\mu$m, or\\

\item The silicates are largely responsible for the 10-$\mu$m continuum level, while also
producing a dip near 16 $\mu$m, which initially looks like a 22-$\mu$m emission
feature in the photometry.
\end{enumerate}
Inspection of the components of the plots in \citet[][e.g. their fig. 8]{li01b}
suggests that the silicates provide a peak at 10 $\mu$m, a continuum dip around
16 $\mu$m and a shoulder near 20 $\mu$m. We have adopted this latter approach, as
also shown in Fig. 8. We discuss these alternatives in Sect. 8.3.
   \begin{figure*}
   \center
   \includegraphics[angle=0,width=15cm]{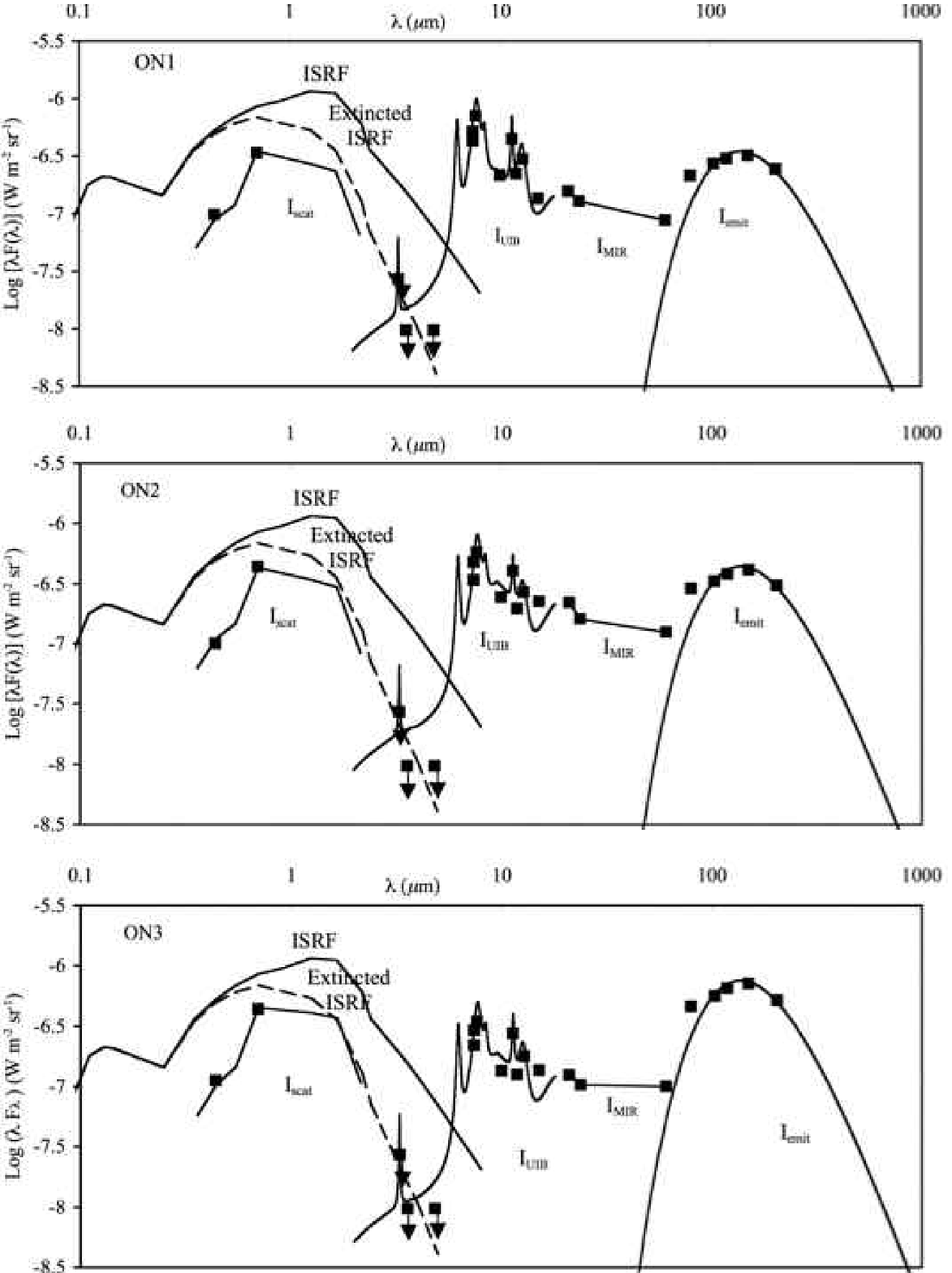}
      \caption{
Spectral Energy Distributions for ON1, ON2 and ON3. The components contributing to the energy balance in the different wavelength ranges are indicated: $I_{\rm{scat}}$ is the scattered UV $-$ NIR light, $I_{\rm{UIB}}$ is the UIB emission, $I_{\rm{MIR}}$ is the VSG MIR contribution and $I_{\rm{emit}}$ is due to the big grains. The total interstellar radiation field (ISRF) in the Solar neighbourhood is also shown both before and after extinction effects (solid and dashed lines respectively). The data points marked on each of the $I_{\rm{scat}}$ lines represents the emission in the $B_J$ and $R$ bands, obtained by plate calibration with the TPN (see Sect. 4).
              }
         \label{Fig10}
   \end{figure*}
\section{ISRF and the energy budget}

The ISRF is the heating source of the dust grains in clouds
lacking internal heating sources. The grains absorb photons in the ultraviolet to near-IR 
range, and subsequently emit photons at mid- to far-IR wavelengths corresponding to their
temparature.

The energy emitted by the large grains, $I_{\rm{emit}}$, was obtained by integrating the
modified blackbody fits from Sect. 5.1 over the wavelength interval 60 $-$ 1000 $\mu$m.
We define $I_{\rm{UIB}}$ and $I_{\rm{MIR}}$ to be the areas under the empirical UIB curve and the
observed 20 $-$ 60 $\mu$m filter fluxes, respectively. Given that the dust albedo is non-
zero, some of the unabsorbed radiation is scattered from the cloud by the dust
particles, producing optical and near-IR surface brightness. Taking the estimated optical
SEDs obtained in Section 4, and integrating them over the wavelength range 0.36 $-$ 2.2 $\mu$m,
we obtained the intensity of scattered radiation, $I_{\rm{scat}}$, for the three positions.
The resultant energy budgets for the three sightlines, ON1, ON2, and ON3, are summarized in Table 5. 
The corresponding spectral energy distributions are shown in Fig. 10. 

The solid ISRF lines in Fig. 10 represent the unattenuated local ISRF according to \citet{mathis83}
(hereafter MMP) as modified by \citet{lehtinen98}: the intensities between  
1.25 $-$ 2.2 $\mu$m were derived using the {\it COBE}/DIRBE all-sky data at the $J, H,$ and $K$ bands,
and the resulting ISRF intensities at these bands were 20 $-$ 50 per cent higher than the
MMP ones, resulting in an integrated  ISRF(0.091 $-$ 8 $\mu$m) intensity of
$20.4 \times 10^{-7}$ W m$^{-2}$ sr$^{-1}$, some 15 per cent larger than the MMP value of 
$17.3 \times 10^{-7}$ W m$^{-2}$ sr$^{-1}$.

However, not all of the total incident ISRF intensity is relevant to the energy
balance of the cloud: only the extincted part of the ISRF contributes. 
Consequently, we have estimated the fraction of the ISRF that suffers extinction in the cloud
by taking into account the wavelength and spatial dependence of the optical depth through
the cloud. The average radial dependence of extinction was derived using the 2MASS extinction
data displayed in Fig. 7. A cloud diameter ($\phi$) of 0.8$^{\circ}$ centered on position ON3 was adopted.
Effective extinction factors were then derived by calculating the term in square brackets in 
equation 5 of \citet{lehtinen98}.
The ISRF spectrum was then reduced in accordance with these extinction
factors and the resulting extincted ISRF intensity,  $I_{\rm{ISRF}}^{\rm{ext}}$,
is shown as a dotted line in Fig. 10 and its total intensity, integrated over 0.091 $-$ 8 $\mu$m, 
at the bottom row of Table 5.

The quantity to be compared with the extincted ISRF intensity is not, however, the integrated surface
brightness at any individual position of the cloud, but the mean surface brightness over the cloud face,
$\overline{I(\lambda)}$, integrated over all wavelengths (for details, see equations 5 and 6 of
 \citealt{lehtinen98}).
In order to account for this, we have approximated G~300.2~$-$~16.8 as a spherical cloud of diameter
0.8$^{\circ}$ (c.f. Fig. 1) and determined from the two ESO/SERC and four {\it IRAS} maps the mean
surface brightness within this boundary. The background level, determined at our reference positions,
 has been subtracted.
The ratios of the mean surface brightness to the surface brightness values at the three ON positions
 are given in Table 6. 
Using these ratios, i.e. ESO/SERC $B_J$ and $R$ for $I_{\rm{scat}}$, {\it IRAS} 12 $\mu$m for
$I_{\rm{UIB}}$, {\it IRAS} 25 and 60 $\mu$m for $I_{\rm{MIR}}$, and  {\it IRAS} 100 $\mu$m for
 $I_{\rm{emit}}$, we
have derived the three estimates for the total mean radiation $\overline{I_{\rm{tot}}}$
(row 6) and the average of the three estimates in row (7). The {\it IRAS} maps are thus being used
to obtain an estimate for the spatial distribution over the cloud face, while the ISOPHOT data
are used to determine its spectral distribution. Within the error limits,
$\overline{I_{\rm{tot}}}$ is seen to be in good agreement with 
the extincted ISRF $(0.091-8\mu\rm{m})$ intensity.

To ensure that this result did not strongly depend on the selected
cloud boundary, we also repeated this procedure using cloud diameters of  0.7$^{\circ}$ and 0.9$^{\circ}$.
For all three boundaries tested,
the ratios  $\overline{I_{\rm{tot}}} / I_{\rm{ISRF}}^{\rm{ext}}$ were obtained. They were
0.99$\pm$0.05, 0.89$\pm$0.05, and 0.88$\pm$0.05 for  0.7$^{\circ}$,  0.8$^{\circ}$, and 0.9$^{\circ}$, respectively.
This suggests that a 0.8$^{\circ}$ boundary is a reasonable
 assumption and that  $\overline{I_{\rm{tot}}}$ and $ I_{\rm{ISRF}}^{\rm{ext}}$
are equal within the error limits. We will discuss the details of the energy budget in Sect. 8.6.

\begin{table}
\center
   \caption[5]{Summary of the energy budget.}
      \label{table5}
\begin{tabular}{llll}
\hline
\multicolumn{ 1}{l}{ISM component and} & \multicolumn{ 3}{c}{Surface Brightness [$10^{-7}$ W m$^{-2}$ sr$^{-1}$]} \\
\multicolumn{ 1}{l}{wavelength integration} & & & \\
\multicolumn{ 1}{l}{range} & ON1 & ON2 & ON3 \\
\hline
Scattered radiation, $I_{\rm{scat}}$ & \multicolumn{ 1}{l}{$4.0\pm 1.0$} & \multicolumn{ 1}{l}{$5.0\pm 1.3$} & \multicolumn{ 1}{l}{$5.6\pm 1.4$} \\

$(0.36-2.2\mu\rm{m})$ & \multicolumn{ 1}{l}{} & \multicolumn{ 1}{l}{} & \multicolumn{ 1}{l}{} \\

UIB emission, $I_{\rm{UIB}}$ & \multicolumn{ 1}{l}{$3.6\pm 0.7$} & \multicolumn{ 1}{l}{$3.6\pm 0.7$} & \multicolumn{ 1}{l}{$2.2\pm 0.4$} \\

$(6-18\mu\rm{m})$ & \multicolumn{ 1}{l}{} & \multicolumn{ 1}{l}{} & \multicolumn{ 1}{l}{} \\

VSG MIR emission, $I_{\rm{MIR}}$ & \multicolumn{ 1}{l}{$1.4\pm 0.3$} & \multicolumn{ 1}{l}{$1.9\pm 0.4$} & \multicolumn{ 1}{l}{$1.2\pm 0.2$} \\

$(21-60\mu\rm{m})$ & \multicolumn{ 1}{l}{} & \multicolumn{ 1}{l}{} & \multicolumn{ 1}{l}{} \\

Large grain emission, $I_{\rm{emit}}$ & \multicolumn{ 1}{l}{$3.6\pm 0.7$} & \multicolumn{ 1}{l}{$4.6\pm 0.9$} & \multicolumn{ 1}{l}{$7.9\pm 1.6$} \\

$(60-1000\mu\rm{m})$ & \multicolumn{ 1}{l}{} & \multicolumn{ 1}{l}{} & \multicolumn{ 1}{l}{} \\

Total radiation, $I_{\rm{tot}}$ & $12.7\pm 2.7$ & $15.1\pm 3.3$ & $16.9\pm 3.6$ \\
(scattered + emitted) & &\\
Estimated $\overline{I_{\rm{tot}}}$ & $10.4 \pm 2.4$ & $10.6 \pm 2.5 $ & $11.5 \pm 2.7 $ \\ 
for $\phi = 0.8^{\circ}$ area        &      &      &      \\
\hline
 \multicolumn{ 2}{l}{Average $\overline{I_{\rm{tot}}}$ for $\phi = 0.8^{\circ}$ area} & \multicolumn{ 2}{c}{10.8 $\pm 2.5$} \\
 \multicolumn{ 2}{l}{Total ISRF, $I_{\rm{ISRF}}(0.091-8\mu\rm{m})^{\mathrm{a}}$} &        \multicolumn{ 2}{c}{20.4} \\

\multicolumn{ 2}{l}{Extincted ISRF, $I_{\rm{ISRF}}^{\rm{ext}}(0.091-8\mu\rm{m})$} &        \multicolumn{ 2}{c}{12.2} \\

\hline
\end{tabular}
\begin{list}{}{}
\item[$^{\mathrm{a}}$] \citet{mathis83} as modified by \citet{lehtinen98}
\end{list}
\end{table}

\begin{table}
\center
   \caption[6]{Surface brightness ratios $\overline{I(\phi = 0.8^{\circ})}$/$I$(ONi). The estimated errors are 10 per cent.\\} 
      \label{table6}
\begin{tabular}{llll}
\hline
\multicolumn{ 1}{l}{ } & ON1 & ON2 & ON3 \\
\hline
\multicolumn{ 1}{l}{Wavelength Band:} & \multicolumn{ 3}{c}{ } \\
$B_J$  & 1.00 & 0.98 & 0.88 \\
$R$  & 0.82 & 0.63 & 0.63 \\
IRAS 12$\mu$m  & 0.65 & 0.66 & 1.29 \\
IRAS 25$\mu$m & 0.80 & 0.56 & 0.86 \\
IRAS 60$\mu$m & 0.83 & 0.58 & 0.47 \\
IRAS 100$\mu$m & 0.89 & 0.69 & 0.46 \\

\hline
\end{tabular}
\end{table}
\section{Far-infrared opacity}

We derive the ratio between the FIR optical depth $\tau_{\rm{em}}(\lambda)$ and the
optical extinction, $\tau_{200} / A(V)$ as well as the average absorption cross
section per H-atom, $\sigma_{\lambda}^{\rm{H}} = \tau(\lambda) / N(H)$. If this
ratio can be assumed to be the same from cloud to cloud, one can estimate the
total cloud masses by using only the observed far-IR fluxes and the dust temperature
derived from these fluxes \citep{hildebrand}.

In the case of optically thin emission ($\tau$($\lambda$) $\ll$ 1) and an isothermal
cloud, the observed surface brightness is \\
\\
$I(\lambda) \approx \tau(\lambda) B(\lambda, T_{\rm{dust}})
\propto \lambda^{-\alpha}  B(\lambda, T_{\rm{dust}})$ \\
\\
We have derived the optical depths for $\alpha=2$ and for the $T_{\rm{dust}}$ values as
given in Table 3. The resulting values of $\tau_{200}$ are listed in Table 7.

\begin{table}
\center
   \caption[7]{Visual extinction, optical depth and average absorption cross-sections
per H-nucleon for ON1, ON2 and ON3. $\tau_{200}$ is calculated under the assumption
that $\tau_{\lambda} \propto \lambda^{-2}$.}
      \label{Table7}
\begin{tabular}{llll}
\hline
           &        ON1 &        ON2 &        ON3 \\
\hline
   $A(V)$ & $1.8\pm 0.2$ & $1.9\pm 0.2$ & $2.9\pm 0.2$ \\

$N(H)$ $[10^{21}$ cm$^{-2}]$ $^{a}$ &       $3.5\pm 0.4$ &       $3.7\pm 0.4$ &       $5.66\pm 0.4$ \\

$I_{200} (\nu)$ [MJy sr$^{-1}$] &       $16.6\pm 3.4$ &       $20.8\pm 4.3$ &       $35.2\pm 7.2$ \\

$\tau_{200}$ $[10^{-4}]$ & $1.9\pm 0.2$ & $2.4\pm 0.2$ & $4.0\pm ^{0.4}_{0.3}$ \\

$I_{200} / A(V)$ [MJy sr$^{-1}$] &        $9.2\pm 2.1$ &       $11.0\pm 2.5$ &       $12.1\pm 2.6$ \\

$\tau_{200} / A(V)$ $[10^{-4}\rm{ mag}^{-1}]$ &   $1.1\pm 0.2$ &   $1.3\pm 0.2$ &   $1.4\pm ^{0.2}_{0.1}$ \\

$\sigma_{200}^{\rm{H}} = \tau_{200} / N(H)$ & $0.6\pm^{0.1}_{0.2}$ & $0.7\pm 0.1$ & $0.7\pm 0.1$ \\
$[10^{-25}\rm{~cm}^{2}\rm{~H~nucleon}^{-1}]$ & & & \\
\hline
\end{tabular}
\begin{list}{}{}
\item[$^{\mathrm{a}}$] The total hydrogen column density $N(H)$ is derived from
$A(V)$ using the $N(H)$ vs. $E(B-V)$ relationship of
\citet{bohlin} and $R(V) = 3.05$.
\end{list}
\end{table}
Using the NIR extinction values derived in Sect. 4 or the equivalent $A(V)$ values as
given in Table 2, we can derive values for $I_{200} / A(V)$ and $\tau_{200} / A(V)$,
as given in Table 7.

We can also use the extinction for estimating the total hydrogen column densities,
$N(H + H_{2})$. As a starting point, we adopt the value of $N(H+ H_{2}) / E(B - V) =
5.8 \times 10^{21} \rm{cm}^{-2} \rm{mag}^{-1}$ for diffuse clouds \citep{bohlin},
together with $R(V) = 3.05$ to obtain $N(H + H_2) / A(V) = 1.90 \times 10^{21} \rm{cm}^{-2}
\rm{mag}^{-1}$. We thus derive for G~300.2~$-$~16.8 the total hydrogen column densities
as shown in Table 7, and from these we calculate the values of $\sigma_{\lambda}^{\rm{H}}$,
as listed in the last line of Table 7.

\section{Discussion}
\subsection{ON1, ON2, ON3 in comparison with diffuse ISM sightlines}
Our three observed ON positions in G~300.2~$-$~16.8 cover a wide range of {\it IRAS} band
ratios, $I_{12 \mu\rm{m}} / I_{25 \mu\rm{m}}$ and $I_{12 \mu\rm{m}} / I_{100 \mu\rm{m}}$,
comparable to almost the whole range of these parameters as observed in a large
sample of high-latitude diffuse and translucent clouds by \citet{laureijs89a}. In this
respect, the three SEDs for ON1, ON2 and ON3 may be considered as representative
templates for diffuse/translucent sight lines in general. The three SEDs and their
differences clearly suggest the presence of at least three MIR$-$FIR emitting dust components,
with the UIBs strongest at ON1, the VSG emission making its largest relative
contribution at ON2, and the large, classical grains dominating the IR emission
spectrum at ON3.

The emission spectra derived from our data may be compared with those of previous
datasets, such as the \citet{dwek} SEDs produced using {\it COBE} data. The Dwek
et al. data were successfully modelled by them using a three-component model. In
their work, PAHs provided the $3.5 - 4.9$-$\mu$m emission in their models, the MIR-range
emission was provided by VSGs, and at wavelengths above 140 $\mu$m, the
emission was dominated by $T \approx 17 - 20$ K graphite and $15 - 18$ K silicate
grains, in general agreement with our observations. Their model featured lower
elemental abundance requirements than preceding work: they only required $\sim$20 per cent of
the total carbon to be in PAHs and $\sim$60$-$70 per cent in graphite, but all of the silicon was
locked in silicate grains. Their SED represented an average over a mixture of several
high-latitude diffuse and translucent ISM clouds. Subsequent work by
\citet{verter} indicated notable cloud-to-cloud variations in the mid-IR emission of high-
latitude translucent clouds, which they attributed to abundance and composition
differences in the grains. The spatial resolution of our {\it ISO} data has permitted us to
detect similar such variations within a single cloud.

   \begin{figure*}
   \center
   \includegraphics[angle=0,width=15.5cm]{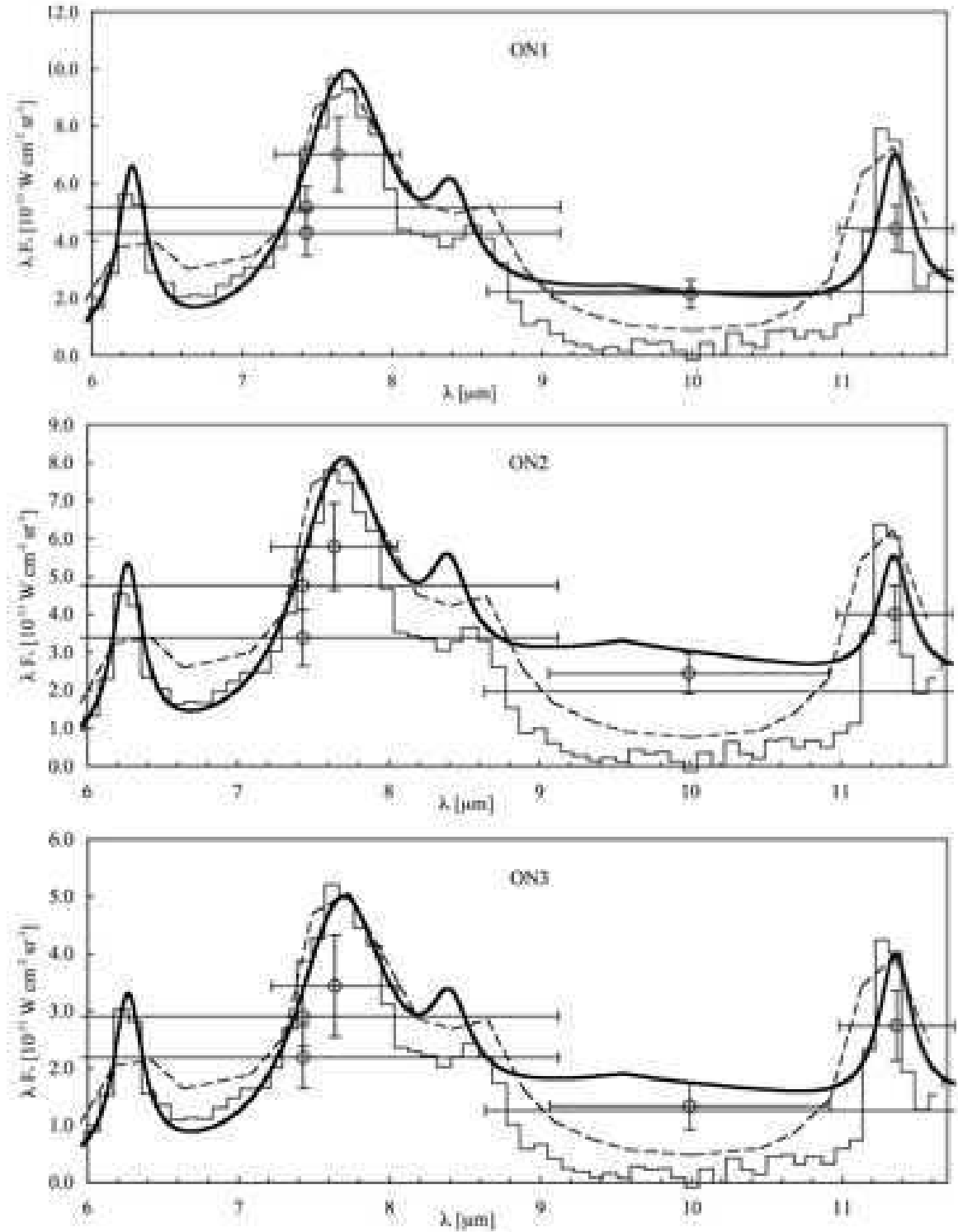}
      \caption{
Comparison of semi-empirical modelling with other Galactic UIB spectra. The thick solid lines are the semi-empirical model fits described in Sect. 5.1 to our ISOPHOT data (open circles). The horizontal bars reflect the effective ISOPHOT band widths, including that of the 11.5-$\mu$m band (beyond the plotted wavelength range), which extends down to $\sim$8.64 $\mu$m. The histogram lines are the average spectrum of the six strongest spectra from \citet{kahanpaa}, multiplied by appropriate scaling factors. The dashed lines are the strongest spectrum of \citet{onaka}, again multiplied by appropriate scaling factors.
              }
         \label{Fig11}
   \end{figure*}

A comparison of the models with the shape of other Galactic UIB spectra from the
literature further illustrates the similarities and differences of the sightlines. Fig. 11
compares the ON1, ON2 and ON3 with the observed average of the six strongest UIB
spectra of \citet{kahanpaa} and the strongest spectrum of
\citet{onaka}. Although the general shapes of the spectra are indeed similar, there are two
notable differences in the $8 - 11$-$\mu$m range between our data and the other spectra.
Firstly, the central position of the 8.6-$\mu$m UIB appears to be at a slightly shorter
wavelength than previously observed. This arises due to our basing of the model on the
\citet{boulanger98} ISOCAM spectrum taken very close to ON1, which appears to
show such properties. Furthermore, the lack of a narrow ISOPHOT filter band
covering this feature means that the properties of the band in our model are not well
constrained, hence, our model should only be taken as an approximation of this
band's wavelength. The second, more significant difference between our data and that
of the \citet{kahanpaa} and \citet{onaka} spectra is that the
continuum near 10 $\mu$m is significantly higher in our spectra. This appears to be a
genuine effect, and will be discussed in Sect. 8.3.
\subsection{UIB properties}
The UIB strengths have already been discussed in L98 and the present study confirms
their results. Results obtained from the semi-empirical
modelling of Sect. 5.1, listed in Table 3, suggest that the spectra can be fitted using
a scalable continuum and a set of six Cauchy profiles, in good agreement with the
CVF data of \citet{boulanger98}. Despite the differences in the continuum-subtracted
band heights between positions ON1 and ON3, the ratio of the 7.7- and 11.3-$\mu$m
UIB heights is similar, indicating that the relative carrier abundances are similar
within errors. The 7.7 $\mu$m/11.3$ \mu$m ratio at ON2 is rather higher than for the
other two positions, but still within 30 per cent. Given the non-uniqueness of the underlying
continuum used, together with the overall uncertainties on the photometry
measurements, the spectrum does not appear significantly different. Furthermore, the
UIB spectra appear to be similar to that seen in other interstellar environments (see
e.g. table 7 of \citealt{li01b} for a comparison).
\subsection{Mid-IR continuum: PAHs vs. silicates}
Our measured 10-$\mu$m continuum level in G~300.2~$-$~16.8 has been confirmed by
\citet{boulanger98}, and also reflects the continuum level in the Polaris cirrus cloud
\citep{boulanger00}. The presence of an underlying continuum between the UIBs
at 10 $\mu$m bears directly on the flux levels at 16 and 20 $\mu$m. On
the basis of our observations, it appears that the 10-$\mu$m continuum level
correlates more closely with the mid-IR emission at $16 - 25$ $\mu$m, rather than the
UIBs (see Fig. 8 and Table 2). Under the
assumption of a \citet{li01b}-type model, there appear to be two alternative explanations for this continuum which does not appear in the \citet{li01b} diffuse ISM results.
The first is enhanced PAH emission through either a greater proportion
of carriers, or enhanced particle emissivity. The second possibility is
an excess of very small silicate particles.

In our physical modelling, the size distributions of silicate and
graphite grains were modified by a factor $a^{\Delta \gamma}$, and the
models preferred a large number of small silicate grains that accounted
for much of the 10-$\mu$m continuum. There was a smooth change to PAH
optical properties around $a \sim$50\AA~for graphite grains. Graphite grains
above the 50\AA\ limit do not provide any significant continuum at
10 $\mu$m. For the PAHs, both the optical properties and the shape of
the size distribution were kept constant. Additional continuum at
10 $\mu$m could therefore only be provided in the models by small
silicate grains. We might also consider the possibility that the increased continuum
level could be produced equally well by small graphite grains ($\la$10\AA). At positions
ON2 and ON3, this would better account for the 16-$\mu$m observations that
were underestimated in the numerical models (Fig. 9). However, (spherical)
graphitic grains smaller than 10 \AA~would be difficult to incorporate on
physical grounds. In the bulk form of graphite, the distance between 
graphene sheets is $\sim$ 3.35 \AA, and the size of a unit cell carbon 
hexagon is $\sim$ 2.46 \AA. These are both already a significant fraction of 
the proposed grain size, and for grains this small, some other assumptions
(e.g. the heat capacities used or the assumption of a continuous energy spectrum)
are also no longer strictly correct. Irrespective of the material forming these grains,
however, the qualitative conclusion about the continuum level should still be valid.

\citet{li01a} presented models with which they addressed the possible
contribution to the diffuse ISM emission spectrum at $\lambda$ $\approx 10 -
30$ $\mu$m attributable to ultrasmall  ($a < 15$ \AA ) silicate dust grains. They also
attempted to place upper limits on the presence of ultrasmall silicates via their
contribution to the UV extinction curve. They estimated that as much as 10 per cent of the
available interstellar silicon could be in ultrasmall silicates without violating any
observational constraints.

In the physical models of Sect. 5.2, the relative amount of silicate was increased
with respect to the amount of graphite. The final row of Table 4 summarizes the
silicon abundance requirements of the models as a multiple of the
\citet{li01b} diffuse ISM requirement. The models all incorporate more silicate by a factor
of $1.5 - 3.3$ than the \citet{li01b} model. Furthermore, under the assumption of
MgFeSiO$_{4}$ grains with a density 3.5 g cm$^{-3}$, the \citet{li01b} models
also have a silicon abundance requirement of $4.85 \times 10^{-5}$, already higher than the
Solar silicon abundance of $3.6 \times 10^{-5}$. The amount of silicate needed could be reduced by
changing the models, e.g. by employing a stronger radiation
field. A separate population of small silicate (or graphite) grains might help to
address the issue. Nevertheless, the availability of the amount of silicon needed
will remain a problem.

Whatever the nature of the particles involved, there is now clear evidence of
underlying emission that cannot be modelled using a first order polynomial, as has
been done previously: there are differences between the baseline levels of the
6.2/7.7/8.6-$\mu$m and 11.3/12.7-$\mu$m groups. Even using a simple semi-empirical
model, it is clear that this continuum appears to peak between 10 and 17 $\mu$m.
Although a modelled silicate continuum has been used in the
semi-empirical fitting in Sect. 5.1, we re-emphasize here that this is only an
example of a possible fit, and does not necessarily
suggest the presence of an excess of ultrasmall silicate grains.
\subsection{The large `classical' grains}
Both the simple empirical fitting and the analytical \citet{li01b}-type model
are able to characterize the broader aspects of the big grain populations in
G~300.2~$-$~16.8.
In all three sightlines, the grain emission in the long-wavelength ISOPHOT
bands can be described by particles radiating at a stable temperature of
$\sim$17.5 K (Table 3). These temperatures are in good agreement with
\citet{verter}, who used DIRBE to
survey and model eight translucent molecular clouds and reported temperatures of
$\sim 16 - 18$K for the big grains. The study of high-latitude clouds by
\citet{delBurgo} indicates a warm dust component with a temperature of
$\sim$17.5 K. Earlier work by \citet{dwek} also suggested that
for wavelengths greater than 140 $\mu$m, their model was dominated by emission
from T$\sim 17 - 20$K graphite and $15 - 18$K silicates.  \citet{li01b} also report
temperatures of $13 - 22$K for a mixture of graphitic \& silicate grains between 0.01 and
0.2$ \mu$m. For the large grains, our numerical modelling adopted the properties
described in \citet{li01b}. In our physical SED models, the graphite
component dominates over the silicate component at wavelengths greater than
$\sim$60 $\mu$m; thus we would expect our $\sigma_{\lambda}^{\rm{H}}$ values to be closer to
that of graphite, which is also the case.

The listed $\sigma_{\lambda}^{\rm{H}}$ estimates for G~300.2~$-$~16.8 were obtained using the
methods of \citet{hildebrand}. \citet{li01b} have calculated absorption cross-sections for a diffuse dust
model. The resulting $\sigma_{\lambda}^{\rm{H}}$ values are given in Table 8 for each of the
grain components, i.e. graphite and silicate particles, as well as for the total mixture. For comparison
with the observations, one must take into account that the observationally determined cross-section
is a weighted sum of the graphite and silicate cross-sections, the weighting factor for the warmer graphite
particles being larger than the one for the cooler silicates. This leads to the situation that in the
\citet{li01b} model, the contributions to the high-latitude emission spectrum by graphite and silicate
grains are almost equal at $\lambda \sim 90 - 150$ $\mu$m (see their fig. 8). The effective cross-section
as determined from the observed SED will thus be between the ``graphite'' and ``silicate'' values (see
\citealt{mezger} for details).  It can be seen that the \citet{li01b} calculated cross-sections are in
reasonable agreement with our estimates for G~300.2~$-$~16.8. 

It appears, however, that the dust population composition towards G~300.2~$-$~16.8 does not simply reflect the classic diffuse ISM picture. In Table 8, we also compare the ratios $\tau_{200}/A(V)$ and $\sigma_{\lambda}^{\rm{H}}=\tau_{200}/N(H)$ in G~300.2~$-$~16.8 with a number of other interstellar clouds and dust models. While the \citet{li01b} $\sigma_{\lambda}^{\rm{H}}$ values are generally consistent with our data, we note that the data in Table 8 suggest that our sightlines are different from the diffuse ISM. While it is unsurprising that the $\tau_{200}/A(V)$ and $\sigma_{\lambda}^{\rm{H}}$ values for the denser environments are much higher than for G~300.2~$-$~16.8, we also note that the \citet{dwek} {\it diffuse} ISM values are almost twice our G~300.2~$-$~16.8 estimates. This suggests that the large grain population contribution is different for our sightlines, and that in general, smaller grains dominate in this region.

\begin{table}
\center
   \caption[8]{Comparison of the 200-$\mu$m optical depth-to-$A(V)$ ratio and average absorption cross-section per H-atom nucleon for a range of ISM environments.}
      \label{table8}
\begin{tabular}{llll}
\hline
    Object & $\tau _{200} / A(V)$ & $\sigma_{200}^{\rm{H}}$  &  Reference \\
& & $= \tau _{200} / N(H)$ & \\
           & $[10^{-4} \rm{mag}^{-1}]$ & $[10^{-25}$ cm$^{2}$ & \\
 & & H nucleon$^{-1}]$ & \\
\hline
G~300.2~$-$~16.8 &            &            &            \\

       ON1 &        $1.1\pm 0.2$ &       $0.6\pm ^{0.1}_{0.2}$ & This work \\

       ON2 &        $1.3\pm 0.2$ &       $0.7\pm 0.1$ &            \\

       ON3 &        $1.4\pm ^{0.2}_{0.1}$ &       $0.7\pm 0.1$ &            \\
\\
Diffuse ISM &        2.5 &        1.4 & Dwek \\
 & & & et al. (1997) \\
\\
Thumbprint &        5.3 & $2.5_{-1.4}^{+3.2}$ & Lehtinen \\
Nebula & & & et al. (1998) \\
(globule) & & & \\
\\
L183 & $3.4-3.8$ &            & Juvela et al. \\
\multicolumn{ 2}{l}{(dense cloud)} & & (2002) \\
\\
Model &            &        2.7 & Hildebrand \\
\multicolumn{ 2}{l}{(molecular clouds)} & & et al.(1983) \\
\\
Model &            & 1.0 (silicates) & Li \& Draine \\
(diffuse ISM) &            & 0.53 (graphite) & (2001b) \\

           &            & 1.53 (total) &            \\
\hline
\end{tabular}
\end{table}
\subsection{UIB/VSG/Large Grain emission variations}
Extinction mapping and modelling of G~300.2~$-$~16.8 by \citet{laureijs89b} suggested
that the variations of the relative emission strengths
in the {\it IRAS} 12, 25, 60 and 100-$\mu$m bands between the centre and edge of the
cloud cannot be solely attributed to ISRF attenuation effects, and they instead proposed that the variations are attributable to real abundance variations of the carrier populations.
Modelling of the IR emission of G~300.2~$-$~16.8 by \citet{bernard93}
also indicated that strong {\it IRAS} colour variations could not be attributed to radiative
transfer effects alone.
Their models instead suggested the presence of a halo of small dust particles and/or
PAHs, with their densities decreasing towards the cloud interiors and giving way to
the dominance of larger grains. This general picture is supported by our
observations: ON3 is near the central part of the cloud, and ON1 and ON2 sample
different halo positions.

More recently, surveys of MIR-FIR/submm dust emission have been conducted using
data from the {\it PRONAOS}/SPM in conjunction with {\it IRAS} and {\it COBE}/DIRBE
data for several sightlines (see \citealt{stepnik} for a review). \citet{bernard99} observed the
high-latitude cirrus cloud MCLD 123.5 +24.9, which exhibited lower large grain
equilibrium temperatures than other similar objects and a MIR emission deficit
measured at 60 $\mu$m. The {\it IRAS} colours of this object already suggested an excess
of cold dust, and the low temperature was attributed to a change in the dust properties
and grain coagulation. Similar results for a molecular filament in the Taurus complex,
coupled with the detection of corresponding submm enhancement, led
\citet{stepnik} to postulate that the effects there can be modelled by VSG-big grain
coagulation into large fluffy aggregates.

Although G~300.2~$-$~16.8 is at high Galactic latitude and does not show a 60-$\mu$m
emission deficit in our data, we conclude that it is not a normal cirrus cloud. It has a relatively
high average large grain temperature of $\sim$17.5 K for its estimated $A(V)$ at all three
observed positions, and appears to exhibit a comparatively large number of small grains.
The clear differences in the relative contributions to the emission spectrum by the UIBs, VSGs and
large grains between the sightlines may be attributable to two scenarios:
\begin{enumerate}
\item Either the relative component population differences were established during the original cloud
formation, and/or \\

\item Ongoing processing may give rise to material exchange between the
populations (e.g. coagulation or fragmentation processes).
\end{enumerate}
Distinguishing between these two possibilities may prove difficult, however, and might require
statistical sampling of a number of clouds to conclusively determine whether
anti-correlations exist between the components. Nevertheless, we note that for ON3,
the increased large grain emission is accompanied by lowered UIB emission, which might
suggest the accretion of the UIB carrier species on to large grains. Given that larger
FIR emissivity appears to be needed for position ON3 (and possibly also for ON2)
than for ON1, grain growth would seem plausible.
\subsection{Energy budget and the ISRF}
Within the error limits, the total surface brightness of G~300.2~$-$~16.8,
integrated from the UV to FIR wavelengths, was seen to be
equal to the mean Solar neighbourhood ISRF intensity($0.091-8\mu\rm{m}$), extincted 
in the cloud (see Sect. 6 and Table 5). 
This supports the scenario in which G~300.2~$-$~16.8 is solely heated by the
external ISRF ($0.091-1.25\mu\rm{m}$). No associated stars are known of in
G~300.2~$-$~16.8 which could supply any noticeable additional heating radiation.

In the physical modelling of the SEDs in Sect. 5.2, the MMP ISRF as 
modified by \citet{lehtinen98} was adopted and a good  agreement was found with the
observed SEDs of positions ON1$-$ON3 not only with respect to their shape, but also
with the absolute flux level. This indicates that the adopted ISRF intensity is of the
correct magnitude to supply the heating power to the cloud. There are, however, other
free parameters in the physical modelling, such as the line-of-sight dust column
density (producing extinction) and the albedo of the grains, which can account for an 
error in the assumed ISRF intensity. The extinction by diffuse dust in the
`envelope« and intercloud region of the Chamaeleon dark cloud complex  may be expected to
attenuate the ISRF incident on G~300.2~$-$~16.8 by $\sim$ 10 -- 15 per cent.
Taking this attenuation into account would increase the $A(V)$ estimates obtained from the
physical modelling by $\sim$ 10 -- 15 per cent, thus bringing them into better agreement with
the directly observed $A(V)$ values.

The direct estimation of the energy balance, as presented in Sect. 6, does not
depend on the detailed radiative transfer and dust modelling. Besides the adopted SED of
the ambient ISRF, only observed quantities are involved, i.e. the extinction profile 
of the cloud and the UV-to-FIR surface brightness averaged over the cloud face.
In this simple form, the approach is valid for clouds with
spherical symmetry. This appears to be a reasonable assumption for G~300.2~$-$~16.8
 (see Figs. 1 and 7). The incident ISRF intensity is not isotropic, but shows
 a strong concentration towards the Galactic equator. For the
cloud emission at mid- and far-infrared wavelengths, this causes no complications
since the cloud emission is isotropic. However, at the UV-to-NIR wavelengths where
the, predominantly forward-directed, scattering dominates, the surface brightness 
of the cloud depends on the viewing direction: a cloud of small-to-moderate
optical depth is brighter when viewed in direction of the Galactic plane as
compared to a viewing direction towards high Galactic latitudes.
In the case of G~300.2~$-$~16.8, the viewing direction is towards an intermediate
Galactic latitude, $b = -16.8^{\circ}$, which means that the {\em effective} ISRF 
intensity for scattering is close to the average ISRF intensity over the sky 
(see \citealt{lehtinen96} for details). 

An interstellar cloud for which both the extinction profile and the UV-to-FIR
surface brightness are measured can be used as a probe of its ambient ISRF
intensity. This way, one could probe the ISRF within
$\sim$ 1 kpc of the Sun in different environments, such as stellar associations or 
clusters with enhanced ISRF. The location of  G~300.2~$-$~16.8 in the Chamaeleon
Cloud Complex is not expected to imply any significant enhancement of the ISRF:
the young stars in Cha I and II clouds are mostly of low luminosity and many
of them are still embedded in their parental cloud.
The analysis of the energy balance of the Thumbprint Nebula, a globule also located
in the Chamaeleon Complex, gave the same result, i.e. that its ambient radiation field
can be well represented with the average Solar neighbourhood ISRF (see \citealt{lehtinen98}). 
\section{Summary}
\begin{enumerate}
\item We have presented multi-wavelength ISOPHOT photometry of three differing
sightlines corresponding to {\it IRAS} 12-, 25- and 100-$\mu$m photometry peaks within the
G~300.2~$-$~16.8 high latitude translucent cloud. The data cover the NIR$-$FIR wavelength
range, and demonstrate the need for at least three ISM components: UIB carriers,
transiently-heated VSGs and large classical grains at thermal equilibrium.\\

\item 2MASS $JHK_{S}$ colour excesses of stars visible through the cloud were used to
construct an extinction map. Star counts in the $B_{J}$ and $I$ bands indicated a
reddening law compatible with the diffuse dust $R(V)$ value.\\

\item The UIB section covered by filter photometry was fitted semi-empirically using a
typical model silicate continuum and six Cauchy curves. An underlying continuum
peaking around $15 - 16$ $\mu$m was found to be necessary. The emission by large
grains was well fitted for all three sightlines with a modified blackbody curve of
$T\sim$17.5 K. This is warmer than expected, but not incompatible with other observations
of diffuse and translucent clouds. Emission was detected in the MIR range, which was
attributed to non-thermal emission by transiently-heated small grains.\\

\item Under the assumption of simple spherical geometry, numerical modelling of the
data, based on the work of \citet{li01b}, was performed. The models were able to fit the
spectra plausibly. The differences in the SEDs cannot be explained by radiation field variations or
radiative transfer effects. Abundance variations are required to explain the differences between
(e.g.) the UIB carrier contributions to the cloud centre (ON3) and halo (ON1) spectra.
Some very small silicate grains appear to be required by the models, with a greater quantity of silicon (and carbon) than is allowed by some current cosmic abundance estimates, but this is true of other existing models as well (e.g. \citealt{li01b}).\\

\item The FIR opacity per nucleon, measured by the quantity
$\sigma_{\lambda}^{\rm{H}}=\tau_{200}/N(H)$ , assumes values in G~300.2~$-$~16.8 which
are lower than some of the recently obtained values for globules and dense molecular clouds.
The estimated values of $\tau_{200}/A(V)$ are also smaller than those obtained from
recent diffuse ISM models, suggesting an enhancement of the small grain population.\\

\item Within errors, the energy of the sum of the optical scattered light and emission from the UIBs,
VSGs and large grains can be supplied by the local ISRF. The relative amounts of the
UIB, VSG and large grain emission vary as expected from the original {\it IRAS}
photometry: UIB emission is strongest in ON1, the mid-IR VSG emission peaks at
ON2, and the large grain emission dominates at ON3. It is not yet clear whether
material is transferred between the three populations, and statistical sampling of a
range of clouds may be required to address this issue in the future. The variations
between the three sightlines appear to be a result of abundance variations of the three
(or more) carrier populations within the cloud, and this may be due to grain
coagulation processes.\\
\end{enumerate}

\section*{Acknowledgments}
We gratefully acknowledge support from the Finnish Academy (grant No. 174854)
and the Magnus Ehrnrooth Foundation (MGR). The authors wish to thank Peter
\'Abrah\'am for advance use of his ZL template spectra prior to their
publication. We thank the referee, Ren\'{e} Laureijs, for very helpful comments. ISOPHOT and the
Data Centre at MPIA, Heidelberg, are funded by the Deutsches Zentrum f\"ur Luft- und
Raumfahrt DLR and the Max-Planck-Gesellschaft. DL is indebted to DLR, Bonn, the
Max-Planck-Society and ESA for supporting the {\it ISO} Active Archive Phase.
This research has made use of NASA's Astrophysics Data System.

\label{lastpage}

\end{document}


\center
\setcounter{table}{0}
\begin{table*}
\center
\caption{Observational ISOPHOT data and accompanying error estimates.}
      \label{Table1}
\begin{tabular}{llllllllllllll}
\hline
    Filter & $\lambda_{C}$ & $\Delta\lambda$ & Aperture & Ref. & TDT Range & $S_\nu$ & $\Delta S_\nu$ & $\sigma_{\rm{INT}}$ & $\sigma_{\rm{EXT}}$ &  $\sigma_{\rm{EXT}}$ /  $\Delta S_\nu$ & Absolute & Filter-to-filter & Adopted Combined \\
 & &  & & & & & & & & & Accuracy & Relative Accuracy & Accuracy \\
 & [$\mu$m]          & [$\mu$m]         &  & & & \multicolumn{2}{|c|}{[MJy sr$^{-1}]$} & \multicolumn{2}{|c|}{[MJy sr$^{-1}]$} &[\%] & [\%]     & [\%] & [\%] \\
 (1) & (2) & (3) & (4) & (5) & (6) & (7) & (8) & (9) & (10) & (11) & (12) & (13) & (14) \\
\hline
   P1\_3.29 &        3.3 &       0.22 & 180'' diam. &        L98 & 27600501$-$27600511 & & $^{\mathrm{a}}$ & & & & & & \\
   P1\_3.60 &       3.59 &          1 & 180'' diam. &        L98 & 27600512$-$27600522 & & $^{\mathrm{a}}$ & & & & & & \\
   P1\_4.85 &       4.86 &       1.55 & 180'' diam. &        L98 & 27100423$-$27100433 & & $^{\mathrm{a}}$ & & & & & & \\
   P1\_7.30 &       7.43 &       3.38 & 99'' diam. &  TW & 70801676$-$70801681 & 6.85$^{\mathrm{b}}$ &       0.81$^{\mathrm{b}}$ &        0.42$^{\mathrm{b}}$ &     0.08$^{\mathrm{b}}$ & 9 &        20 &         10 & 19 \\
   P1\_7.30 &       7.43 &       3.38 & 180'' diam. &        L98 & 21700651$-$21700656 & 4.76$^{\mathrm{b}}$ &       0.96$^{\mathrm{b}}$ &       0.01$^{\mathrm{b}}$ &      0.06$^{\mathrm{b}}$ & 6 &        20 &         10 & 16 \\
    P1\_7.7 &       7.64 &       0.84 & 180'' diam. &        L98 & 21700434$-$21700439 & 6.18$^{\mathrm{b}}$ &       1.28$^{\mathrm{b}}$ &        0.04$^{\mathrm{b}}$ &    0.13$^{\mathrm{b}}$ & 10 &        20 &         10 & 21 \\
     P1\_10 &       9.99 &       1.86 & 180'' diam. &        L98 & 21700457$-$21700467 & 12.49$^{\mathrm{b}}$ &      0.66$^{\mathrm{b}}$ &        0.03$^{\mathrm{b}}$ &        0.09$^{\mathrm{b}}$ & 14 &        20 &         10 & 24 \\
   P1\_11.3 &      11.36 &       0.77 & 180'' diam. &        L98 & 21700440$-$21700450 & 15.31$^{\mathrm{b}}$ &      1.37$^{\mathrm{b}}$ &        0.09$^{\mathrm{b}}$ &        0.12$^{\mathrm{b}}$ & 9 &        20 &         10 & 19 \\
   P1\_11.5 &      11.89 &       6.51 & 180'' diam. &  TW & 70801301$-$70801311 & 17.29$^{\mathrm{b}}$ &      0.67$^{\mathrm{b}}$ &       0.16$^{\mathrm{b}}$ &       0.17$^{\mathrm{b}}$ & 25 &        20 &         10 & 35 \\
   P1\_12.8 &      12.83 &       2.33 & 180'' diam. &        L98 & 21700479$-$21700489 & 16.45$^{\mathrm{b}}$ &      1.06$^{\mathrm{b}}$ &        0.02$^{\mathrm{b}}$ &        0.09$^{\mathrm{b}}$ & 8 &        20 &         10 & 18 \\
     P1\_16 &      15.14 &       2.86 & 180'' diam. &        L98 & 21700468$-$21700478 & 20.43$^{\mathrm{b}}$ &      0.80$^{\mathrm{b}}$ &        0.09$^{\mathrm{b}}$ &        0.12$^{\mathrm{b}}$ & 14 &        20 &         10 & 24 \\
     P2\_20 &      21.08 &       9.43 & 180'' diam. &  TW & 70801312$-$70801322 & 22.60$^{\mathrm{b}}$ &      1.22$^{\mathrm{b}}$ &       0.08$^{\mathrm{b}}$ &       0.24$^{\mathrm{b}}$ & 20 &        20 &         10 & 30 \\
     P2\_25 &      23.81 &       9.18 & 180'' diam. &  TW & 70801327$-$70801333 & 19.89$^{\mathrm{b}}$ &      0.78$^{\mathrm{b}}$ &       0.09$^{\mathrm{b}}$ &       0.17$^{\mathrm{b}}$ & 22 &        20 &         10 & 32 \\
     C1\_60 &       60.8 &       23.9 & 135''$\times$135'' &  TW & 70801340$-$70801344 & 8.81 &      2.05 &        0.12 &       0.16 & 8 &        25 &         20 & 33 \\
     C1\_70 &       80.1 &       49.5 & 135''$\times$135'' &  TW & 70801346$-$70801351 & 18.02 &      8.35 &        0.12 &        0.40 & 5 &        25 &         20 & 30 \\
    C1\_100 &      103.5 &       43.6 & 135''$\times$135'' &  TW & 70801334$-$70801339 & 21.08 &      13.59 &       0.07 &         0.20 & 2 &        25 &         20 & 27 \\
    C2\_120 &        119 &       47.3 & 181''$\times$181'' &  TW & 70801352$-$70801357 & 29.30 &      16.17 &       0.10 &        0.30 & 2 &        20 &         10 & 22 (12)$^{\mathrm{c}}$ \\
    C2\_135 &        150 &       82.5 & 181''$\times$181'' &  TW & 70801358$-$70801363 & 39.22 &      22.71 &       0.07 &        0.54 & 2 &        20 &         10 & 22 (12)$^{\mathrm{c}}$ \\
    C2\_200 &      204.6 &       67.3 & 181''$\times$181'' &  TW & 70801364$-$70801369 & 40.91 &      24.04 &       0.19 &        0.48 & 2 &        20 &        10 & 22 (12)$^{\mathrm{c}}$ \\
\hline
\end{tabular}
\begin{list}{}{}
\item[Column descriptions:]{(1) Filter name; 
(2) Central wavelength, $\lambda_{C}$; 
(3) Filter width, $\Delta\lambda$; 
(4) Aperture size and shape; ``diam.'' denotes the diameter of a circular aperture, `135''$\times$135''' and `181''$\times$181''' denote a square aperture; 
(5) Reference. ``L98'' is Lemke et al. (1998), ``TW'' denotes this work; 
(6) TDTs of observations; 
(7) Mean total sky surface brightness, $S_\nu$, averaged over the three ON positions; 
(8) Mean (ON $-$ REF) surface brightness, $\Delta S_\nu$, averaged over the three ON positions; 
(9) Mean PIA-propagated internal error, $\sigma_{\rm{INT}}$, as described in Sect. 2.4; 
(10) Mean external error, $\sigma_{\rm{EXT}}$, as described in Sect. 2.4; 
(11) $\sigma_{\rm{EXT}}$ as a fraction of $\Delta S_\nu$; 
(12) Absolute accuracy (see Klaas et al. 1994); 
(13) Filter-to-filter relative calibration accuracy; 
(14) Adopted combined accuracy. For the ZL-calibrated PHT-P data, this is taken to be $\sigma_{\rm{EXT}}$ + Filter-to-filter relative calibration accuracy. For the PHT-C data, in the absence of an independent (ZL) calibration curve, this is taken to be $\sigma_{\rm{EXT}}$ + absolute accuracy.}
\item[$^{\mathrm{a}}$]{Upper limit only (see L98).}
\item[$^{\mathrm{b}}$]{Values listed for the P1 and P2 bands are after ZL calibration.}
\item[$^{\mathrm{c}}$]{The filter-to-filter relative calibration accuracy values listed for the C2 data were used to produce the error estimates in parentheses. These were used when estimating uncertainties on FIR temperatures (see Sect. 5.1 and Table 3). These smaller values are justified for this purpose, as these three sets of measurements all use the same detector configuration and only differ in the filter used, and are thus not susceptible to cross-detector calibration uncertainties. The quoted values are based on an estimate of upper limits on remaining uncertainties due to responsivity variations.}
\end{list}
\end{table*}